\documentclass[twocolumn]{article}
\usepackage{graphicx,multicol,fancyhdr,authblk,lastpage,hyperref,amssymb,amsmath,mathtools,tfrupee,algorithm,tikz,algpseudocode,rotating,multirow,amsthm,subcaption,url,balance}
\usepackage[capitalise]{cleveref}
\newtheorem{assumption}{Assumption}

\newcommand{\Probability}{\text{I\kern-0.15em P}}

\fancypagestyle{plain}
{
    \fancyhf{}
    
    \rfoot{\textit{\scriptsize Page \thepage\ of \pageref{LastPage}}}
}

\algdef{SE}
[CLASS]
{Class}
{EndClass}
[1]
{\textbf{class} \textsc{#1}}
{\textbf{end class}}

\title{\huge \textbf{Indian Kidney Exchange Program: A Game Theoretic Perspective}}
\author[1]{\large Arghya Bandyopadhyay\footnote{Department of Computer Science and Engineering, National Institute of Technology, Durgapur, West Bengal, India; Email:\href{mailto:arghyab13bga@gmail.com}{arghyab13bga@gmail.com}}\hspace{0.2cm}and Sajal Mukhopadhyay\footnote{Department of Computer Science and Engineering, National Institute of Technology, Durgapur, West Bengal, India; Email:\href{mailto:sajal@cse.nitdgp.ac.in}{sajal@cse.nitdgp.ac.in}}}
\begin{document}
\thispagestyle{plain}
\maketitle
\begin{abstract}
    \emph{We propose a way in which Kidney exchange can be feasibly, economically and efficiently implemented in Indian medical space, named as Indian Kidney Exchange Program(IKEP) along with Indian specific influence on compatibility and final outcomes. Kidney exchange is a boon for those suffering from renal kidney failure and do have a donor with an incompatible kidney (compatible kidney also encouraged for better matches). In such situations the patient, donor pair is matched to another patient, donor pair having the same problem and are compatible to each other. Hospitals put up their patient-donor data. Using the biological data, compatibility scores(or weights) are generated and preferences are formed accordingly. Indian influence on weights, modify the compatibility scores generated and hence, the preferences. The pairs are then allocated using game theoretic matching algorithms for markets without money.}
    \flushleft{\textbf{Keywords: Kidney exchange, kidney paired donation, donor-recipient pair, living donor kidney transplant.}}
\end{abstract}
\section{Introduction}
 We hereby, propose a centralized, matching algorithm and a framework for a national kidney exchange program in India, named as \textbf{Indian kidney exchange program(IKEP)} motivated from Singh N P et al. 2016\cite{28} which discusses the challenges and future of kidney transplantation in India and states that kidney exchanges programs(KEPs) can be beneficial in Indian kidney transplantation market. Lecture 10 of Tim Roughgarden\cite{40} provides a brief idea on kidney exchange paradigm and various mechanism designs deployed. 
 
 \subsection{Overview}
The patients who are suffering from ESKD and have a donor\footnote{The donor should be eligible to donate to the patient under the Indian act of Organ donation 1994.} are supposed to see for the hospitals associated with IKEP\footnote{We expect all the hospitals to get themselves associated to IKEP, which would help in getting a much more thicker pool.}. The patient will then report to the nearest hospital and get themselves enrolled into the IKEP pool of patient-donor pairs. After successful registration of patient-donor pair in IKEP program, preferences are generated based on the information provided by them (here, the patients and the donors, in realistic sense, are strategic in nature and hence, the generated preferences will be treated in strategic setting) and then, the patients and the donors are matched by a mechanism designed specifically for markets without money (mechanism design without money is a subtopic of game theory\cite{41}).

\begin{figure}[h]
\centering
\includegraphics[width=8cm]{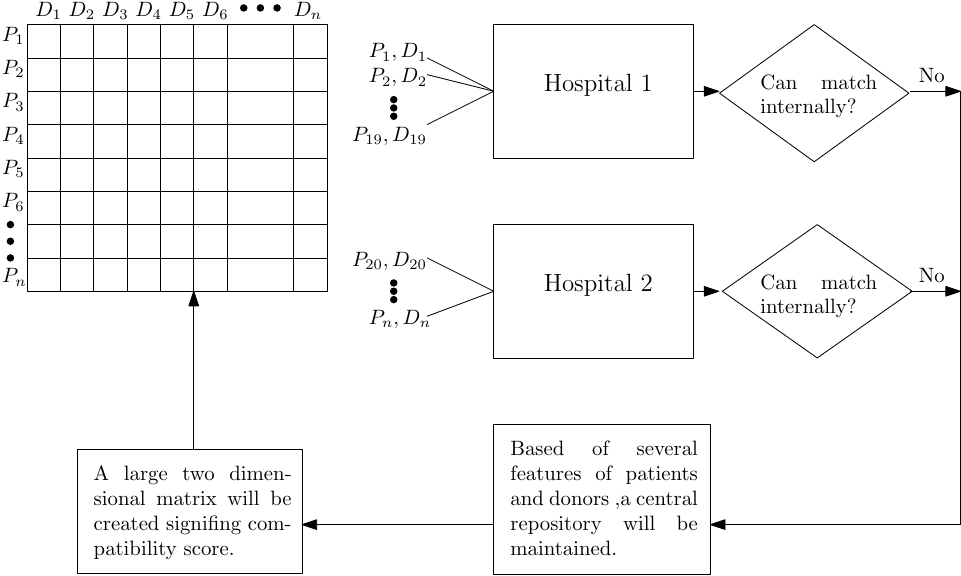}
\caption{Overview of IKEP}
\label{IKEA_overview}
\end{figure}

All the patients, along with their donors, approach the respective hospitals for enrolling themselves in IKEP. Although not promoted\footnote{Because, more number of pairs in the IKEP will lead to increase in the $\mathbf{card}($pairs$)$ getting matched to their preferred donor.}, the hospitals try to match pairs internally and enroll its remaining, unmatched pairs in IKEP. As soon as the pairs are enrolled, a central repository is created based on several features of patients and donors as discussed in the \cref{prerequisites}. After that, the compatibility score\footnote{Compatibility scores and weights are used interchangeably throughout this paper.} of all the patients with each and every donor participating in the IKEP is calculated and stored in a $n \times n$ matrix. The calculations of compatibility scores/weights are discussed in the \cref{weightCalculations}. \cref{IKEA_overview} describes the overview of IKEP where $P_i$ and $D_i$ denotes patient and donor respectively.

From the matrix generated, IKEP analyzes the top most preference of the pairs on the donors. And these preferences are used to generate stable matches between pairs and donors.

In the remaining parts of paper, \cref{currentSituationInIndia} gives us an idea of where India stands in terms of kidney exchange, transplantation and various other prevailing issues. \cref{IKEP_background} provides a basic understanding of kidney exchange and various other technical or related terms. Now in  \cref{systemModel} describes the system model of IKEP and \cref{IKEP_algorithms}, present the algorithmic implementation of the mechanism proposed. We then paint a picture of how Indian influence on weights(Societal acceptance score) affects the compatibility matrix generated to form a preference list in \cref{IndianPrespective}.

After that, in \cref{effectsEFAndEIKEPA} we discuss the effects of Edge Filter, how it works and cons of using the normal IKEPA as described in \cref{IKEP_algorithms}. To solve the problems described, we come up with an Enhanced version of IKEPA i.e, E-IKEPA, \cref{enhancedIKEPA} states the same.

In \cref{prioritizingPairs}, we describe the way in which, the preferences are made strict and non-ambiguous. \cref{IKEP_deployment} speaks about the way in which the cycles proposed by IKEP could be deployed in a distributed but simultaneous manner. After that, in \cref{simulationUsingIndianData}, we simulate IKEP run on Indian Data. \cref{limitationsOfIkep} states the limitations of IKEP and how they could be solved. \cref{futureScope} and \cref{conclusion} provides insights for future work and concludes the paper. 

\cref{currentSituationInIndia} gives an idea of the current situation of KEPs in India and things related to that. and \cref{IKEP_background} provides necessary background details about kidney exchange and biological terms used in this paper.

\subsection{Prerequisites}
\label{prerequisites}
Before enrolling a patient into the Indian kidney exchange program, the hospitals are required to keep the following details of the donor and patients respectively:
\begin{enumerate}
    \item \textbf{Age.} The age of the donor matters\cite{30}, It has been observed that patients getting a kidney from an age group of 18-34 years has 62\% of graft survival and 59\% for kidneys from age group 35-49 and so on.\footnote{Referred from  \href{https://portal.kidneyregistry.org/compatible_pairs.php}{National  Kidney Registry USA}.}
    \item \textbf{Blood Group.} (Ex: A, B, AB, O) It is collected to decide ABO type compatibility discussed in \cref{biologicalConstraint}.
    \item \textbf{List of Human leukocyte antigens -A, -B, -DR antigens} The list helps in calculating the compatibility score of a patient from pair 1 with the a donor from pair 2.
    \item \textbf{Kidney Size} The size of kidney also plays a role in increasing the survival of the patient after the graft.
    \item \textbf{Panel Reactive Antibodies(PRA) percentage.} The PRA is crucial in deciding the waiting time of a particular patient. The PRA points to the level of sensitization of the patient. Patients with higher PRA, face more difficulty in getting matched.
    \item \textbf{Pin code.} If the pin-code is found same, it increases the acceptance and feasibility of the exchange and reduces with increase in the distance between two pin-codes.
    \item \textbf{Has ever donated a kidney.} This data will help in prioritizing the patients in the pool.
    \item \textbf{Societal preference.} It is basically a ranked list of acceptable societal distributions\footnote{The term societal distribution is self explanatory. Every society is distributed by factors like, Religion, caste and creed.}, with most preferred being the first element and least preferred being the last element. The societal distributions, not present in the list are assumed unacceptable. If the societal preference is found $\emptyset$, it is assumed that the patient is indifferent to all the kidneys present in the pool in terms of societal distributions.
    \item \textbf{Societal distribution of donor.} The societal distribution of donor is noted down in order to check the acceptance of his/her kidney by any patients according to their societal preference.
    \item \textbf{Distance from dialysis center.} This parameter is used to allocate the marginal priority which is used for breaking the ties described in \cref{prioritizingPairs}.
    \item \textbf{Economic slab.} The patient falling in lower economic slab are marginally prioritized over the higher economic slabs.
\end{enumerate}

\subsection{Timeline}
\label{timelineOfIKEP}
The timeline of IKEP basically consists three steps which will provide a clear picture of the entire process.
\subsubsection{Submitting medical data} 
Hospitals submit their patient donor pair medical details to enroll them as soon as the patients approach their hospitals.
\subsubsection{Matching and exchange proposals} 
Matching \cref{algo:ikep_main} is carried out after a particular interval and exchange cycles are proposed to the hospitals for final verdicts.
\subsubsection{Offer reviews and transplants}
As soon as the matching cycles are proposed by the algorithm, a cross match test is performed by the central testing center for IKEP and based of the results final matching results are presented to the concerned hospitals. The hospitals review the exchanges offered. On approval, the the next step of transplantation is taken, else the pairs remain in the exchange pool and get considered for next match.

\section{Current Situation in India}
\label{currentSituationInIndia}
In this section, we paint the scenario of medical, legal, and various other frameworks revolving around kidney or any other organ transplantation. We also show a glimpse of burden of ESKD on the medical system and current situation of handling the burden. We also talk about various issues which provide hindrance towards KEP(s).

It has been noticed that kidney failures are becoming a leading cause for many deaths. Number of deaths happening due to Chronic Kidney Disease(CKD) in India raised from 0.6 million in 1990 to 1.18 million in 2016 \cite{33}. According to an article by \href{https://www.indiatoday.in/education-today/gk-current-affairs/story/world-kidney-day-hypertension-and-diabetes-two-major-causes-of-kidney-diseases-1477841-2019-03-14}{India Today} there are approximately 1,00,000 people who are diagnosed with End-stage kidney disease(ESKD) every year. And shows the burden of CKD in Indian medical space. And one of the best method to cope up with this burden is Kidney transplantation. Many patients rely on dialysis(two to three times a week) for there life, which is much more costlier than both in terms of money, time and effort. And more importantly quality of life degrades considerably. 

Kidney exchange(KE) is essential for India since its absence leads to a considerable $\mathbf{card}($pairs$)$ get matched with incompatible kidneys which in turn needs to be treated with immunosuppressants\footnote{Immunosuppressants are those medical drugs given to the patients in order to stop there body from rejecting the new kidney transplanted. } in large amounts.\cite{37} The use of immunosuppressants add to a lifelong monthly expenditure on medicines which is approximately near about \rupee 10,000 to \rupee 15,000. This expense can also be reduced through KEPs by getting more compatible kidneys through them. A considerable amount of work\footnote{Although there are some people working in this domain.} needs to be done in the direction of adaptation of kidney exchange in India to reduce fatalities mentioned above.

One good thing about the kidneys are that every person on this planet are are born with two kidneys(excluding those in case born with one). And can live perfectly fine with any one of them. Therefore, people can donate one of their kidneys to there loved ones, making themselves a living donor. And with the advent of living donors in the field of kidney transplant, comes a situation in which living donor is incompatible to donate to the patient, he/she wants to donate to. This problem is addressed by kidney exchange.

There was an estimate in 2018 stating that in India, there were near about 175,000 patients dependent on chronic dialysis. Which turns out to be 129 per million population.\cite{34} According to a \href{https://www.narayanahealth.org/blog/kidney-transplants-in-india/}{blog by \textbf{Narayana Health}} in the year 2019, there were 200,000 recipient awaiting kidneys, and only 15,000 donors willing to donate kidneys. Moreover, according to a calculation by Ministry of Health, out of a range of 2-3 lakhs transplant requirements, only 6000 are getting accomplished. The burden of fatalities on India is larger than other low-middle income countries. 

And in order to combat the burden, kidney transplants act as a boon. Kidney transplants are far inexpensive to dialysis. According to a study\cite{31}, 60\% of the dialysis patients resides more than 50km away from Hemodialysis(HD) station, and nearly 25\% resides more than 100km away. A huge amount of money in transportation itself.\footnote{At an average, in India there are 2 dialysis per week.} Total cost per dialysis is approximately \rupee4148(\$64), which accounts for \rupee8296(\$128) per week\footnote{In India, there are people working at a salary of \rupee8296(\$128) per month.}\cite{36}. There are portals for deceased kidney transplantation, but the chances are very thin\footnote{The number of deceased kidneys extracted are very less in number because of less time span between the arrival of kidney to the selection of recipient, large recipient waiting pool and less number of deceased kidneys being eligible to be transplanted.}. 

\subsection{Relevant Improvements Towards Kidney Exchange Programs In India} 
The factor behind the success of single centered KEP depends on the formation central registry of the incompatible pairs and the literacy about KEPs\footnote{Most of the people in India don't have information regarding KEPs. This is a new paradigm, which needs to be implemented massively to reduce a significant part of deaths. Discussed more specifically in \cref{literacy}} (Kute VB et al 2017\cite{29}). Single center KE were performed by Pahwa et al\cite{20}, Waigankar et al\cite{22}, Jha et al \cite{26}, Kute et al\cite{23}, Kute et al\cite{24} for the durations 2006-2011, 2008-2011, 2010-2013, 2000-2012, 2013 respectively. Utkarsh Verma et al 2020\cite{35} analysed how multiple registries of KEPs in India work and how can they be brought together and perform matching in coalition with individual rationality constraints. 

\subsection{National Organ Transplant Program} 
\label{nationalOrganTransplanationProgram}
National organ transplant program is an Indian program run by Directorate General of Health Services under the Ministry of Health \& Family Welfare, to procure deceased donor kidneys, maintain EKPD patients waiting list and allot kidneys accordingly. It's objective is to improve the ease of transplantation by promoting deceased organ donations. NOTP also protects vulnerable patients from organ trafficking. NOTTO : National Organ \& Tissue Transplant Organisation act as an apex organisation for coordinating and procuring deceased kidney and providing it to the wait-listed candidates.

\subsection{Legal Framework in India}
\label{legalFrameworkInIndia}
The fist act governing the transplantation in India is \href{https://legislative.gov.in/sites/default/files/A1994-42.pdf}{\textbf{Transplantation of Human Organ Act(THO) 1994}}. 
\begin{itemize}
    \item It was passed in order to regulate the organ procurement, storage and transplantation. 
    \item It also prohibited commercial trading of human organs.
\end{itemize}

 In 2011, an amendment was passed named as \textbf{Transplantation of Human Organ(Amendment) Act(THO) 2011}. 
 \begin{itemize}
     \item It legalizes \emph{swap donations}, i.e., kidney exchanges are legalized by this act.
     \item It specifies that patients are only allowed to receive an organ(here, kidney) from their first relatives only. In situations when there is no first relative, the patient and the donor need to get permission from government appointed authorization committee, which finds out the genuineness of the donor.
     \item The Kidney exchanges need to be approved by the concerned authorities before conducting them.
 \end{itemize}

\subsection{Issues}
There are several issues, prevailing in India which are hindering the progress of kidney exchange paradigm in India.
\subsubsection{Literacy}
\label{literacy}
Literacy of KEPs in India is very less. There is also a shortage of living donor due to the fear of consequences of not having a kidney which actually is none. Recently, Indian government faced a huge resistance from rural part of the country in order to get themselves vaccinated. Hence getting more and more people to participating in organ donation, even to there loved ones will be a big checkpoint.

\subsubsection{Restrictions on NDD}
According to the legal restrictions discussed above in \cref{legalFrameworkInIndia}, non directed donors are not allowed which in turn prohibits chains. And there prohibitions will lead to accumulation of hard to match pairs. Chains have been acting as a boon for those pairs in countries like UK, USA, etc.

\subsubsection{Poor Infrastructure}
India spends approximately 1.5\% of its GDP in healthcare sector. For a vast and populated country like India, the mentioned expenditure is very less. As a result of this, there are significantly huge cardinality of hospitals unmaintained. The government run hospitals do not have adequate  surgical facilities. Infrastructure would add to the resistance towards the implementation of IKEP.

\subsubsection{Abundance of uninsured patients} Since, India is a developing nation, there is a significant amount of poverty prevailing in the country. The poverty stricken population will obviously not consider purchasing insurances. And hence, there would be a section of population, tending towards dying at home than to enroll in a dialysis initially and then in IKEP. 

\subsubsection{Ethical issues} Anonymity is one of the most important ethical issues in KEPs. Some of them encourage anonymity\cite{7,4,12,10,17,13,6,14,15} contrasting others.

\subsection{Existing Kidney Registry}

There are basically two significant kidney registries in India.
\subsubsection{ASTRA} Apex Swap Transplant Registry\footnote{\href{http://www.apexkidneyfoundation.org/astra/}{http://www.apexkidneyfoundation.org/astra/}} - Apex Kidney Foundation, situated in Mumbai performs simultaneous KEPs\footnote{Discussed earlier, way in which procurement and transplantation is carried our simultaneously} in India. It is a paired kidney registry, established in 2011.
\subsubsection{IKDRC} Dr.  HL Trivedi Institute of Trans-plantation  Sciences  (IKDRC-ITS)\footnote{\href{http://www.ikdrc-its.org/}{http://www.ikdrc-its.org/}},  Ahmedabad, India carried out 10 pair non-simultaneous cycles\footnote{Non Simultaneous cycles are not allowed in many countries because of the risk of patient not getting a kidney after providing one by its incompatible donor} in the year 2020 \cite{38} and a simultaneous 10 pair cycle in 2013 \cite{21}. They have significantly contributed towards spreading awareness about KEP in India. Form January 2000 to July 2016, the same hospital carried out 3616 live donor renal transplants and 561 deceased kidney transplants, out to which 300 were kidney exchanges\cite{29}. 

\section{Background and Definitions} 
\label{IKEP_background}
Now, as we know the current situation of kidney exchange programs in India and things revolving around it, We now provide a basic understanding of what kidney exchange is, it's history, and define various terminologies related to it.

\subsection{History}
The paradigm of kidney exchange was first suggested and published by Felix Rapaport in Rapaport FT et al. 1986\cite{2}and was first, successfully carried out in South Korea\cite{3}. After this, several KEPs were carried out in Switzerland\cite{3}, USA\cite{5} and England\cite{18}. The paradigm of KEPs were popularized by Roth et al. (2004)\cite{8}, Roth et al. (2005)\cite{9}, and Roth et al. (2007)\cite{11}

\subsection{Kidney Exchange}
A problem in which at-least two patient-donor\footnote{Sometimes there are patients with their loved ones willing to donate a kidney but are incompetent to receive that kidney.} pairs\footnote{For a patient to participate and benefit from a KE program, it is compulsory to have a donor.}, who are incapable of exchanging kidney among themselves due to incompatibilities discussed briefly in upcoming subsections. But, are compatible among each other. Suppose there are two pairs $(P_1,D_1)$, $(P_2,D_2)$ and $D_1$ can't donate her kidney to $P_1$ rather can donate to $P_2$. Similarly can $D_2$ donate her kidney to $P_1$. And in this way, the exchange happens in a 2-way exchange. KE can accumulate n-way exchanges\footnote{Many countries don't allow more than 3-way exchange. Where as some countries like Germany don't even allow Kidney Exchange.}, but comes with huge costs. Various ways of matching are used for matching the pairs and forming cycles and chains.

\subsection{Cycles and Chains.}
Cycles are formed when there are finite pairs in the pool which can exchange kidneys among themselves. All the matched pairs are brought into a same place and the exchange(operation) takes place simultaneously.

There are some people who voluntarily come up in order to donate a kidney, but in turn don't have any patient along with them. And hence introduce a free-undirected kidney in the system. These people are known as Non-directed Donors(NDD)\footnote{India, France, Poland, Greece, Switzerland are some of the many countries that don't allow live donation of organs}. Presence of these donors lead to the formation of chains. Forming a chain, results into exchanges taking place at different times. Formation of chains reduces logistical constraints, and hence involve large exchanges.

\subsection{Biological constraints}
\label{biologicalConstraint} 
The basic factors for matching a pair to another are dependent on blood group, tissue compatibility and sensitization.
\begin{enumerate}
    \item \textbf{ABO compatibility} Everyone is born with a particular blood group $\in$ [A,B,AB,O]. And in order to receive a kidney form a donor, donor's blood group should not have antigens, absent in the patient's blood.\footnote{Donor with blood group A, can give his/her kidney to a patient with blood group A, AB. Blood group B, can give his/her kidney to a patient with blood group B, AB. Blood group O, can give his/her kidney to a patient with blood group A, B, AB, O.}
    
    \begin{figure}[h]
    \centering
    \includegraphics[width=8cm]{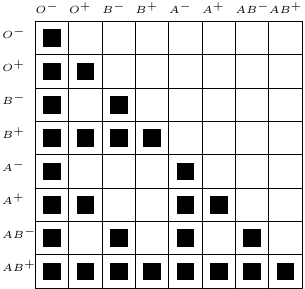}
    \caption{Abo Compatibility}
    \label{aboCompatibilityMatrix}
    \end{figure}
    
    The black squares denote possibility of exchange.

    \item \textbf{Tissue compatibility} Donors and patients have human leukocyte antigens (HLA) which is also of a concern. It has been observed that considering HLA matching has showed better graft survival. These are inherited from parents genetically. And in order to receive a kidney from the donor, the patient shouldn't have antibodies to the HLA carried by donor. HLA consists 1 or 2 antigens depended on Parents having same of different HLA. HLAs are also generated at the time of pregnancy.

    \item \textbf{Sensitization} Some patients do have highly developed antibodies which are reluctant and highly sensitive to the addition of any antigens in the body. Patients with high sensitization denotes that extra caution should be taken for matching and is extremely difficult to get a kidney for her. The basis of measurement is Panel Reactive Antibodies(PRA). Higher the amount of PRA points towards lesser probability of getting a match for the patient. 
\end{enumerate}

\subsection{Crossmatches}
\label{crossmatchDefinition}Even after being ABO and HLA compatible, there is a physical crossmatch\footnote{For all the patient-donor pair who have been matched virtually, blood samples are received from the donor and patient before the transplant and tests are carried upon.} carried out in order to assure that transplant is possible. A positive crossmatch points towards incompatibility and vice-versa.
There are basically two types of crossmatches referred to, in this journal, Virtual cross-match and Serological cross-match.
\begin{enumerate}
    \item \textbf{Virtual Cross-match}
    are the test done based on theoretical knowledge on blood groups, HLA antibodies and various other factors. Based on the results of virtual cross-matches, matching algorithms are carried out. The matches and exchanges proposed by the algorithms are then passed through Serological cross-match test.

    \item \textbf{Serological cross-match} is the test in which donor's blood is added to the patient plasma or serum and is checked for agglutination. If found, signifies incompatibility of donor and patient.
\end{enumerate}

\subsection{Pair types.} 
\label{pairTypes}
Patient-donor pairs are classified based on the difficulty to be matched. The easy to match pairs are those which are highly demanded and vice versa. Mostly ABO compatible pairs(AB-A, AB-B, A-O, AB-O, B-O) are highly demanded as compared to the ABO incompatible pairs(A-AB, B-AB, O-A, O-B, O-AB), because ABO compatible pairs offer more acceptable kidneys than they themselves require.\footnote{Eg., B-O pair offers a O blood group kidney, which is highly demanded for O blood group patients whereas wants an B blood group kidney in return which is considerably easy to get. Moreover pairs like B-O don't generally appear in the pool since they are ABO compatible whereas O-B does accumulate.} Another reason for a pair to be a hard to match pair is high sensitization of the patient.

\subsection{Matching}
Whenever a suitable pool is formed, matching algorithms are carried out in order to match maximum number of pairs. There are basically two types of matching in kidney exchange paradigm.
\subsubsection{Static Matching}
The static matching conducts matching on the pairs present in the pool at a particular timestamp. Pairs participating in successful exchanges are removed from the pool and rest remain in the pool for next time. The probability of getting better matches on waiting is not taken in to account.
\subsubsection{Dynamic Matching}
\label{dynamicMatchingBackground}
Dynamic matching takes probability of getting a match for the pairs in the pool by waiting in to consideration.

It accounts for arrival rates, probability of compatible, incompatible pairs opting out of IKEP after not getting matched or get matched in an easier manner, outside. 

There are different policies used in matching pairs in the market using dynamic matching.
\begin{itemize}
    \item \textbf{Greedy policy.} In this policy, the pairs are matched as soon as they arrive in the market.
    \item \textbf{Batching policy.} In this policy, the pairs are matched in batches at a particular time interval. The time interval varies from programs to programs. Programs wait from one day to weekly to monthly.
\end{itemize}

There is a trade off between the possibility of getting a match for the pair by waiting and taking the risk of losing the life of the patient.

Now, in the next section, we head towards the system model proposed for Indian Kidney Exchange Program, which is a game theoretic approach towards solving the kidney exchange problem.

\section{System Model}
\label{systemModel}

We hereby, propose the system model of IKEP which is a variant of Top-trading cycle used for House allocation problem by Shapley and Scarf et al. 1974 \cite{1}. Roth et al.(2004)\cite{8} proposes that kidney exchange problem can be understood as a special case of house allocation problem. 

The basic idea was to replace the residents who have preferences over houses available with the patients, who have preferences over the kidneys available. And patients point towards their most preferred kidney and it has been observed that, a cycle could definitely be found(always). These cycles are marked out and removed from the graph. 

After the removal of the patients participating in the previous cycles, the remaining patients reassign their most preferred kidney over the remaining kidneys and cycles are reformed and eliminated. It also speaks about the inclusion of chains and there formation at times when no cycles can be found. But in India, Non directed donors(NDD) are prohibited which in turn prohibits formation and carrying out chains. 

The entire preference list is generated based on the IKEP Compatibility Graph. It is a weighted directed graph(E,V,W). Let $V_i,V_j \in V$ be two vertices, and $w_{ij}\in W$ be the weight for the edge $e_{ij} \in E$ from $V_i$ to $V_j$ representing how compatible\footnote{The rate of graft survival of the transplantation.} is the donor of $V_i$ to the patient of $V_j$. $w_{ij}=0$, means that donor of $V_i$ can't donate to the patient of $V_j$.

Before every algorithm run, the weights are recalculated and edges are redrawn to and from the previous members to the new pair. It is assumed that the pair arrive consecutively.

\subsection{Weight calculation.} 
\label{weightCalculations}
IKEP calculate the weights($w_{ij}$) on the basis of HLA matching score($h_{ij}$), PRA percentage, Blood group($b_{ij}$) and some others. The following are the general influences on the weights. \cref{IndianPrespective} states the influence on the weights in Indian perspective.
\subsubsection{Age score($a_{ij}$)}
\label{sm:ageScore}

Age score is the measurement of age based compatibility of donor of node $i$ and patient of node $j$. Since the improvement of graft survival rate and quality is linear with the reduction in age difference, the calculation of weight need to be linear too.

    \begin{equation}
        \label{eq:ageDiff}
         Age_d = Age_i - Age_j
    \end{equation}
    
    Where, $Age_d$, $Age_i$ and $Age_j$ are difference in ages, age of the donor of node $i$ and age of the patient of node $j$ respectively. 
    
    \begin{equation}
        \label{eq:ageScore}
        a_{ij} = 
        \begin{dcases}
            V_a,              & \text{if } Age_d<0\\
            V_a-\alpha_a \times Age_d,      & \text{if } Age_d\leq D_a\\
            0,      & \text{otherwise}
        \end{dcases}
    \end{equation}

    Where $V_a$ is the maximum value generated by \cref{eq:ageScore} and $\alpha_a \in (0,1)$ such that, the increase in $Age_d$ linearly generates $a{ij} \in [0, V_a]$. $D_a$ is a limit to the age difference, beyond which the age score becomes 0. 
    
    We take $V_a$ to be 6, $\alpha_a$ to be 0.15 and $D_a$ to be 40. The \cref{eq:ageScore} has been inspired by from Utkarsh Verma et al. 2020\cite{35}.
    
\subsubsection{ABO score($b_{ij}$)}
\label{sm:aboScore}

ABO score is the measurement of Blood group type compatibility of donor of node $i$ and patient of node $j$.
    \begin{equation}
        \label{eq:aboScore}
        b_{ij} = 
        \begin{dcases}
        V_b,      & \text{if } compatible(B_i,B_j)\\
        0,      & \text{otherwise}
    \end{dcases}
    \end{equation}
    
    Where $V_b$ is the maximum value that could be assigned to $b_{ij}$ in \cref{eq:aboScore}. $V_b \in (0,\infty)$. We take $V_b$ as 6.
    
\subsubsection{HLA score($h_{ij}$)}
\label{sm:hlaScore}

HLA score depends on the cardinality of human leukocyte antibodies matches\footnote{Here, 2 entries each of 3 human leukocyte antibodies types are taken into consideration for the weight calculation and scoring, namely - HLA -A, -B, -DR antigens.} among the donor of the node $i$ and patient of the node $j$. Basu et al. 2008\cite{16} states the benefits of considering HLAs in matching of Kidneys. It speaks about how HLA mismatch calculations can lead to better graft survival.
    \begin{equation}
        \label{eq:matchedHLA}
        m_{ij} = H_i \cap H_j
    \end{equation}
    
    where $H_i$ and $H_j$ is the set of human leukocyte antibodies of the donor of node $i$ and the patient of node $j$ respectively. And, $m_{ij}$ is the set of matched antibodies.
    \begin{equation}
        \label{eq:hlaScore}
        h_{ij} = |m_{ij}|
    \end{equation}
    
    $H^*$ is the number of HLAs being recorded at the time of data input. In this paper, we consider the length of HLA to be 6. Larger the $H^*$, larger is will be the distribution of weights and the preferences will be more strict.
    
\subsubsection{Kidney size score($k_{ij}$)}
\label{sm:kidneyScore}

Kidney score is the measurement of Kidney size compatibility of donor of node $i$ and patient of node $j$.
    \begin{equation}
        \label{eq:kidneyDiff}
         K_d = K_i - K_j
    \end{equation}
    
    Where, $K_d$, $K_i$ and $K_j$ are difference in kidney sizes, kidney size of the donor of node $i$ and kidney size of the patient of node $j$ respectively.
    
    \begin{equation}
        \label{eq:kidneyScore}
        k_{ij} = 
        \begin{dcases}
            V_k-K_d,      & \text{if } K_d\leq D_k\\
            0,      & \text{otherwise}
        \end{dcases}
    \end{equation}
    
    Where $V_k$, is the maximum value generated by \cref{eq:kidneyScore} which is reduced by $K_d$. $D_k$ is the difference in the kidney sizes, beyond which kidney score reduces to zero. We take $V_k$ as 6 and $D_k$ as 3.
    
\subsubsection{Pincode score($p_{ij}$)}
\label{sm:pinScore}

Pincode score is a measure of the logistical compatibility. In the pincodes, the first digit signify zones, the second digit signify sub-zones. Although not of much significance, is calculated by the \cref{algo:pinScore}.
    \begin{equation}
        \label{eq:pincodeScore}
        p_{ij} = pinScore(Pin_i,Pin_j)
    \end{equation}
    where, $Pin_i$ and $Pin_j$ be the pincode of node $i$ and $j$ respectively.

\subsubsection{General Perspective Weight($gw_{ij}$)} \label{sm:generalWeight} 

General perspective weight is the summation of all the above mentioned scores. This, for edge $e_{ij}$ is calculated as:

\begin{equation}
    \label{eq:generalWeight}
    gw_{ij} = h_{ij}+b_{ij}+k_{ij}+a_{ij}+p_{ij}
\end{equation}

The values of the constants $V_a$, $V_b$, $V_k$, $H^*$ and $V_p$ should be kept identical in order to give equal significance to all the scores. And to assign different priorities to the scores, the values assigned to the constants need to be altered accordingly\footnote{The constants with greater values than the remaining ones, would eventually become more significant in compatibility calculation.}. 

The best way to select an identical value for the constants is to decide the number of HLAs to be considered for matching i.e. $H^*$. When decided, set the value of $V_a$, $V_b$, $V_k$ and $V_p$ to $H^*$. And eventually the other depending constants.

\section{Proposed Algorithm}
\label{IKEP_algorithms}
After stating the system model, in this section, we provide the detailed algorithmic representation of IKEP. We also put up an example to visualize the algorithm being run on a randomly generated data.

The \cref{algo:ikep_main} is the main/core of IKEP allocation system which returns the list of exchanges proposed by the algorithm. A loop runs until and unless all the vertices\footnote{The Vertex class is defined in the \cref{classDefinitions}.} are removed or no more cycles can be derived from the remaining vertices and their respective preferences. In the loop, firstly it generates the compatibility graph using the \cref{algo:genCompatibilityMatrix}. After that a queue is formed from the participating vertices for the particular iteration.

\begin{algorithm}[H]
  \caption{IKEPA}\label{algo:ikep_main}
  \begin{algorithmic}[1]
    \Procedure{IKEPA}{$V,A,R$}\Comment{Generates cycles for IKEP Exchanges}
        \State \textbf{initialize} $C= new\ ArrayList<>()$
      \While{$V \neq null$}\Comment{End, if no vertices left}
      
        \State $W = $GenCompatibilityMatrix$(V)$
        \State $Q = $ArrayToQueue$(V)$
        \State $c = $FindCycle$(W,V.size,Q)$
        \If{$c \neq null$}
            \State RemoveCycle$(C,V,W)$
            \State $C.$add$(c)$
        \Else 
            \State $Break$
        \EndIf
      \EndWhile\label{ikepWhile}
      \State \textbf{return} $C$\Comment{List of cycles}
    \EndProcedure
  \end{algorithmic}
\end{algorithm}

\cref{algo:assignInitialPriority} assigns a priority based on the factors discussed at the starting of each algorithm run. It takes one argument, the set of vertices in the pair. If isInitial member variable of any vertex is true, the \cref{algo:assignInitialPriority} sets the value of waiting time score(WTScore) to zero, add all other scores to calculate priority and set isInitial to false. However, if isInitial is false, WTScore is incremented by a unit and then the priority is recalculated.

The \cref{algo:genCompatibilityMatrix} is responsible for generating the compatibility matrix as stated earlier. For every node, it determines the blood group compatibility and if found compatible, HLA compatibility is calculated. The \cref{IndianPrespective} redefines this algorithm in \cref{algo:genCompatibilityMatrix_extended} with an added feature.

On getting the score, if found greater than 0, further score calculations are done. Or else weight for that particular edge is assigned 0. The remaining scores are basically used for breaking ties and forming the preference list more strict.

\begin{algorithm}[H]
  \caption{Generate Compatibility Matrix}\label{algo:genCompatibilityMatrix}
  \begin{algorithmic}[1]
    \Require{$V_{1} \dots V_{N}$}
    \Procedure{genCompatibilityMatrix}{$V$}\Comment{Generate Compatibility Graph}
    \State \textbf{Initialize} $W(e_{ij}) = 0$, for all $e_{ij} \in E$
    \ForAll{$i\in V$}
    \ForAll{$j\in V$}
    \State $b_{ij} = aboScore(DB_i,B_j)$
    \If{$b_{ij}=0$}
        \State $w_{ij} = 0$
        \State \textbf{continue}
    \EndIf
    \State $h_{ij} = hlaScore(DH_i,H_j)$
    \If{$h_{ij}=0$}
        \State $w_{ij} = 0$
        \State \textbf{continue}
    \EndIf
    \State $a_{ij} = ageScore(DAge_i,Age_j)$
    \State $k_{ij} = kidneyScore(DK_i,K_j)$
    \State $p_{ij} = pinScore(DPin_i,Pin_j)$
    \State $w_{ij} = h_{ij}+b_{ij}+k_{ij}+a_{ij}+p_{ij}$
    \EndFor
    \EndFor
    \State \Return{$W$}
    \EndProcedure
  \end{algorithmic}
\end{algorithm}

After that a First In First Out queue is generated for all the vertices left in the pool. Hereafter a cycle is found out using the \cref{algo:ikep_findCycle}, and stored in a global list of cycles. The vertices/nodes participating in the cycle found is then removed. 

Let there be 5 pairs $(P_1,D_1) \dots (P_5,D_5)$. And the characteristics described in the \cref{tab:exampleCharacteristicTable}.

\begin{sidewaystable}
\caption{Example participant details.}
\label{tab:exampleCharacteristicTable}
\vspace{2 mm}
\begin{tabular}{c c c c c c c c} \hline \\
Name  & Blood Group & HLA &Age &Kidney Size &PinCode &Initial Priority &Top Preference \\ \hline
$P_1$ & $A^-$  & A1,B8,DR10,A3,B14,DR17 &45 &11 &496001 &\multirow{2}{*}{$5$} &\multirow{2}{*}{$D_2$}\\
$D_1$ & $B^+$  & A2,B7,DR11,A10,B16,DR8 &30 &12 &496001\\
$P_2$ & $AB^+$ & A1,B8,DR10,A2,B7,DR11 &25 &10 &490020 &\multirow{2}{*}{$2$} &\multirow{2}{*}{$D_3$}\\
$D_2$ & $O^-$  & A1,B8,DR10,A10,B16,DR8 &55 &11.5 &496001\\
$P_3$ & $A^-$  & A1,B8,DR17,A10,B16,DR8 &67 &12 &496001 & \multirow{2}{*}{$1$} &\multirow{2}{*}{$D_2$} \\
$D_3$ & $A^+$  & A1,B8,DR17,A10,B16,DR8 &25 &11 &496001\\
$P_4$ & $AB^+$ & A1,B8,DR17,A10,B16,DR8 &30 &10 &496001 & \multirow{2}{*}{$6$} &\multirow{2}{*}{$D_3$} \\
$D_4$ & $B^-$  & A1,B8,DR17,A10,B16,DR8 &27 &12 &496001 \\
$P_5$ & $A^+$  & A1,B8,DR17,A10,B16,DR8 &60 &11.5 &496001 & \multirow{2}{*}{$10$} &\multirow{2}{*}{$D_3$}\\
$D_5$ & $O^+$  & A1,B8,DR17,A10,B16,DR8 &55 &11.5 &496001\\ \hline
\end{tabular}
\end{sidewaystable}

\begin{figure}[h]
\centering
\includegraphics{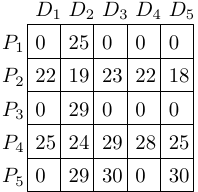}
\caption{Compatibility Matrix}
\label{IKEA_compatibilityMatrix}
\end{figure}

According to the algorithm stated, the compatibility matrix W generated is shown in \cref{IKEA_compatibilityMatrix}. Each value signify the ease of acceptance of donor's kidney by patient's body. Higher value signifies, higher acceptance.

\begin{figure}[H]
\centering
\includegraphics[width=8cm]{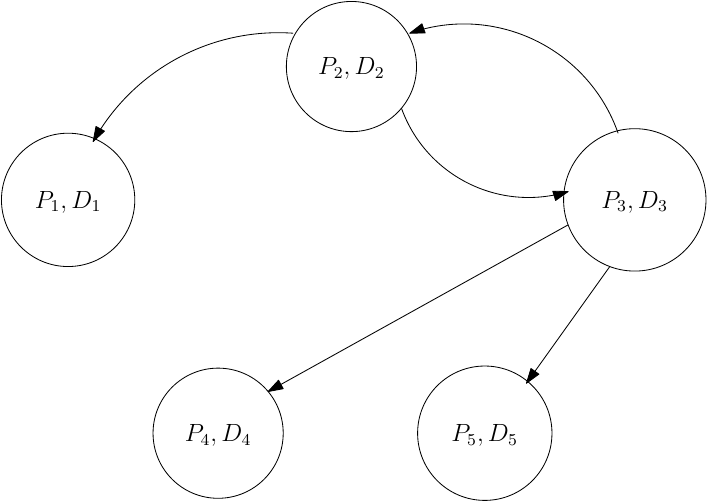}
\caption{Compatibility Graph}
\label{IKEA_compatibilityGraph}
\end{figure}

The graph representation of the compatibility matrix is shown in \cref{IKEA_compatibilityGraph}

Let us assume the priorities\footnote{It is actually assigned and modified with time according to various time dependent characteristics like results of previous algorithm runs, ease to match a pair and many others described in \cref{prioritizingPairs}} assigned to the patients be as described in the \cref{tab:exampleCharacteristicTable}. And the top preferences are derived from the compatibility matrix.

The Graph class defined in \cref{algo:ikep_graphClass}, has member variables $n$ and $adj$ holds the number of nodes in the graph and the list of adjacent nodes respectively. The member method \emph{addEdge} adds directed edges from source node $i$ to destination node $j$. The constructor initializes the graph by taking the number of nodes and initialize the list of adjacent nodes.

\begin{algorithm}[H]
  \caption{Find Cycles}\label{algo:ikep_findCycle}
  \begin{algorithmic}[1]
    \Procedure{findCycle}{$W,S,Q,V$}\Comment{Finds out the cycle including the top prioritized patients.}
        \State $g = $genGraphMaxRowVal$(W,S,V)$
        \While{$Q.$isNotEmpty$()$}
        \State $v = Q.$poll$()$
        \State $c = $dfsCycle$(g,v)$
        \If{$c \neq null$}
        \State \textbf{return} $c$\Comment{proposed cycle}
        \EndIf
        \EndWhile
        \State \textbf{return} $null$\Comment{no cycles possible}
    \EndProcedure
  \end{algorithmic}
\end{algorithm}

IKEP finds the cycles with the help of adjacency matrix that is being passed to the algorithm. The cell with the highest value assigned in a particular row is considered as the most preferred kidney for that particular patient. In case of a tie, the tie breaking is done using priorities assigned to the patients as described in \cref{prioritizingPairs}. The tie breaking algorithm is inculcated in the \emph{max} function.

The cycles are formed using \cref{algo:ikep_dfsCycle} which is a modified version of DFS algorithm. 

\begin{algorithm}[H]
  \caption{Generate Graph Using Max Row Value}\label{algo:ikep_genGraphMaxRowVal}
  \begin{algorithmic}[1]
    \Procedure{genGraphMaxRowVal}{$W,S,V$}\Comment{Generates graph based on highest prioritized kidney for a patient.}
        \State \textbf{initialize} $G= $new Graph$(S)$
        \ForAll{$w_i \in W$}\Comment{$w_i$ is $i^{th}$ row in adjacency matrix}
        \State $i,j = $max$(w)$ \Comment{if max is non-distinct, use pair priority to break ties.}
        \If{max$(w)\geq F^*$}
        \label{algoline:edgeFilter}
        \State $G.$addEdge$(i,V[j])$
        \Else
        \State $G.$addEdge$(i,null)$
        \EndIf
        \EndFor
        \State \textbf{return} $G$\Comment{graph with max row values}
    \EndProcedure
  \end{algorithmic}
\end{algorithm}
$F^*$ is the filter constant which removes the edges having weights less than itself. IKEP adds all the vertices it comes across, in to a hash-set and keeps on checking if the node is repeated. As soon as a repeated node is found, it signifies a cycle. Hence the hash-set is converted into a list and returned back. 

One of the greatest drawbacks of implementation of Top trading cycle is that it could generate such long cycles which would not be feasible to carry out practically. Hence, limiting the length of the cycles is necessary. This is the point, where $L_{max}$ comes in. \cref{algo:ikep_dfsCycle} also keeps on checking the maximum length ($L_{max}$) of the cycle. As soon as the length gets exceeded, the particular cycle is discarded. Different $L_{max}$ leads to different outcomes of IKEP.

\begin{algorithm}[H]
  \caption{DFS Cycle}\label{algo:ikep_dfsCycle}
  \begin{algorithmic}[1]
    \Procedure{dfsCycle}{$g,v$}\Comment{Finds out the cycle starting from v.}
        \State \textbf{initialize} $hs = $new HashSet$<>()$
        \State \textbf{initialize} $start = v$
        \State $hs.$add$(v)$
        \State $v= g.adj.$get$(v).$first$()$
        \While{$v \neq null$}
        \If{$start = v$}
        \State \textbf{return} $hs.$toList()\Comment{cycle emerging from v}
        \ElsIf{$hs.$length$()\geq L_{max}$}
        \State \textbf{return} $null$\Comment{max length exceeded}
        \ElsIf{$!\ hs.$contains$(v)$}
        \State $hs.$add$(v)$
        \Else
        \State $start^*= $hs.$indexOf$(v)\Comment{start index of new cycle found}
        \State $end^*= $hs.$size$()\Comment{end index of new cycle found}
        \State \textbf{return} $hs.$sublist($start^*,end^*$)\Comment{returns the new cycle found}
        \EndIf
        \State $v = g.adj.$get$(v).$first$()$
        \EndWhile
        \State \textbf{return} $null$\Comment{no cycles found}
    \EndProcedure
  \end{algorithmic}
\end{algorithm}

\begin{figure}[h]
\centering
\includegraphics[width=8cm]{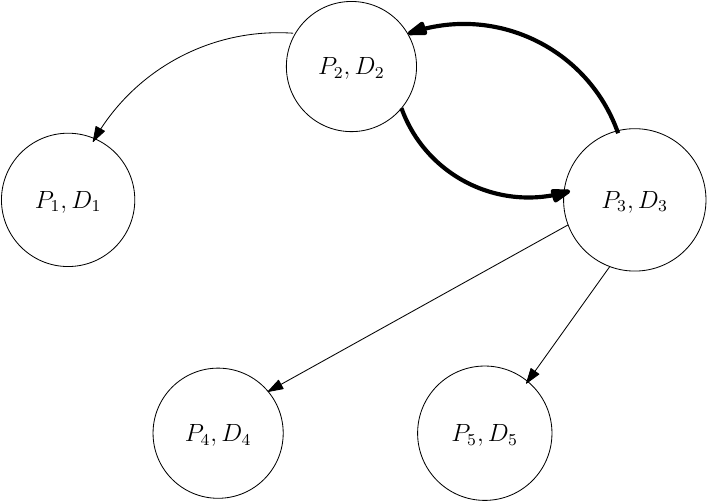}
\caption{Proposed Cycle in First Iteration}
\label{IKEA_compatibilityGraphProposedCycle}
\end{figure}

\cref{IKEA_compatibilityGraphProposedCycle} shows the proposed cycle from the compatibility graph in \cref{IKEA_compatibilityGraph}. 

All the pairs participating in the proposed cycles are removed out of the pool and a new graph\footnote{The Graph class and its members are defined in the \cref{classDefinitions}.} is created based on the top most preference of the patient out of the remaining pairs in the pool and edges are added accordingly.

The \cref{algo:ikep_removeCycle} takes a cycle and list of nodes as arguments and removes the vertices present in the cycle from the list of vertices.
\begin{algorithm}[H]
  \caption{Remove Cycle}
  \label{algo:ikep_removeCycle}
  \begin{algorithmic}[1]
    \Procedure{removeCycle}{$C,V,W$}\Comment{removes out the cycle.}
        \ForAll{$v \in C$}
        \State $index = V.$remove$(v)$
        \ForAll{$W_{ij} \in W$}
        \If{$i=index \lor j=index$}
        \State $W_{ij}=-1$
        \EndIf
        \EndFor
        \EndFor
    \EndProcedure
  \end{algorithmic}
\end{algorithm}

\begin{figure}[H]
\centering
\includegraphics[width=8cm]{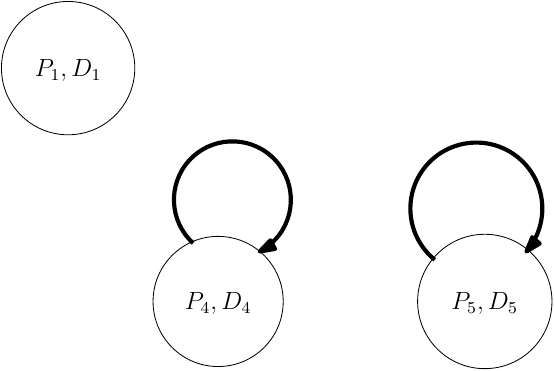}
\caption{\centering Proposed Cycles in Second \& Third Iteration}
\label{IKEA_newCompatibilityGraph}
\end{figure}

After removal of the cycle proposed, \cref{IKEA_newCompatibilityGraph} shows the new graph with redefined edges. And in the next 2 iteration the pairs $(P_4,D_4)$ and $(P_5,D_5)$ are removed. The pair $(P_1,D_1)$ will have to wait for next time.

Now, we state the algorithms to calculate different scores required to weight the edges.

\begin{algorithm}[H]
  \caption{Calculate ABO Score}\label{algo:aboScore}
  \begin{algorithmic}[1]
    \Procedure{aboScore}{$B_i,B_j$}\Comment{Calculate ABO score}
    \If{isCompatible$(B_j, B_i)$}
        \State \Return{$V_b$}
    \Else 
        \State \Return{$0$}
    \EndIf
    \EndProcedure
  \end{algorithmic}
\end{algorithm}

$V_b$ is defined in \cref{sm:aboScore}. The function isCompatible in the \cref{algo:aboScore} returns a Boolean value based on the blood group compatibility and RH factor. It returns the score based on \cref{aboCompatibilityMatrix} in the \cref{biologicalConstraint}. IKEP takes a matrix on compatibility and decides accordingly.

The \cref{algo:ageScore} alters the weight of the edges by presenting a score which is linearly based on the difference between the ages of the patients and donors respectively. Where $V_a,\ \alpha_a,$ and $D_a$ have been described in \cref{sm:ageScore}.
\begin{algorithm}[H]
  \caption{Calculate Age Score}\label{algo:ageScore}
  \begin{algorithmic}[1]
    \Procedure{ageScore}{$Age_i,Age_j$}\Comment{Calculate Age compatibility score}
    \State \textbf{Initialize} $Age_d = Age_i-Age_j$
    \If{$Age_d<0$}
        \State \Return{$V_a$}
    \ElsIf{$Age_d\leq D_a$}
        \State \Return{$V_a-\alpha_a \times Age_d$}
    \Else 
        \State \Return{$0$}
    \EndIf
    \EndProcedure
  \end{algorithmic}
\end{algorithm}

The \cref{algo:hlaScore} calculates score by finding the number of matching human leukocyte antigens. It takes $H^*$ HLAs of a patient and a donor respectively and match them as stated in \cref{sm:hlaScore}.
\begin{algorithm}[H]
  \caption{Calculate HLA Score}\label{algo:hlaScore}
  \begin{algorithmic}[1]
    \Procedure{hlaScore}{$H_i,H_j$}\Comment{Calculate HLA compatibility score}
    \State \textbf{Initialize} $n = 0$
    \ForAll{$h \in H_i$}
        \ForAll{$i \in H_j$}
            \If{$h=i$}
                \State $n= n+1$
            \EndIf
        \EndFor
    \EndFor
    \If{$n \geq H^*$}
        \State \Return{$n$}
    \Else 
        \State \Return{$0$}
    \EndIf
    \EndProcedure
  \end{algorithmic}
\end{algorithm}

The \cref{algo:kidneyScore} prioritizes the exchanges by the compatibility of the kidneys based on the sizes. Lesser the difference in sizes, more likely are they to get matched. The values $V_k$ and $D_k$ are defined in \cref{sm:kidneyScore}.

\begin{algorithm}[H]
  \caption{Calculate Kidney Score}\label{algo:kidneyScore}
  \begin{algorithmic}[1]
    \Procedure{kidneyScore}{$K_i,K_j$}\Comment{Calculate Kidney compatibility score}
    \State \textbf{Initialize} $K_d = |K_i-K_j|$
    \If{$K_d \leq D_k$}
        \State \Return{$V_k-K_d$}
    \Else
        \State \Return{0}
    \EndIf
    \EndProcedure
  \end{algorithmic}
\end{algorithm}

The \cref{algo:pinScore} prioritizes the exchanges by the ease of logistical feasibility of the exchanges. Lesser the distance between the pairs, more likely are they to get matched.

\begin{algorithm}[H]
  \caption{Calculate Pincode Score}\label{algo:pinScore}
  \begin{algorithmic}[1]
    \Procedure{pinCode}{$Pin_i,Pin_j$}\Comment{Calculate Pincode compatibility score}
    \If{$Pin_i$=$Pin_j$}
    \State  \Return{$V_p$}
    \ElsIf{city$(Pin_i)$=city$(Pin_j)$}
    \State  \Return{$V_p-D_p$}
    \ElsIf{subzone$(Pin_i)$=subzone$(Pin_j)$}
    \State  \Return{$V_p-\alpha_{p1}\times D_p$}
    \ElsIf{zone$(Pin_i)$=zone$(Pin_j)$}
    \State  \Return{$V_p-\alpha_{p2}\times D_p$}
    \Else
     \State  \Return{0}
    \EndIf 
    \EndProcedure
  \end{algorithmic}
\end{algorithm}

Where $V_p$ is the maximum value \cref{algo:pinScore} can return and $D_p$ is the rate at which $pinScore$ is deteriorated in successive \emph{if} statements. $D_p$ is multiplied by $\alpha_{p1}\in (1,\infty)$ in order to regulate the rate of deterioration, when sub-zones are identical. Similarly $D_p$ is multiplied by $\alpha_{p2} \in (\alpha_{p1},\infty)$  in order to regulate the rate of deterioration, when zones are identical. We take $V_p$ as 6, $D_p$ as 1, $\alpha_{p1}$ as 2 and $\alpha_{p2}$ as 3.

In the next section, we are going to propose an Indian Influence on the weight calculated till now and briefly visualize it's effect on overall weight. The influence can be visualized more clearly in, \cref{simulationUsingIndianData}.

\section{Indian Influence On Weights}
\label{IndianPrespective}
The compatibility scores/weights generated in \cref{weightCalculations} are biologically driven and do portray the ideal preferences on kidneys. But in India, \textbf{Societal distribution compatibility} act as a significant Indian influence on weights. In this section, we mechanise an algorithm to take it into account and modify the preferences previously generated and redefine \cref{algo:genCompatibilityMatrix}. 

In India, religion, caste and other forms of diversity plays a significant role in Indian healthcare system. Roger P Worthington et al. 2011\cite{19} describes, how Indian medicare and healthcare systems are being influenced and effected by the religious and cultural aspects. India is culturally and societally highly diverse in nature. These poses a huge challenge in forming a nationalized kidney exchange paradigm in India.

Let patient $P_i$ belong to a societal distribution\footnote{Societal distributions in terms of religious, caste, creed and any other form of diversity.} represented by $sd_i$ and let patient $P_j$ belong to $sd_j$. Where, $sd_i \neq sd_j$ and $sd_i, sd_j \in SD$. Where, $SD$ is the set of $n_{sd}$ societal distribution prevailing in a society such that, 
\[SD= \{sd_1,sd_2,\dots sd_{n_{sd}}\}\]

Let us assume, that there is a prevailing hate or societal discrimination for $sd_j$ from $sd_i$. There \textbf{could} be an incidence where $P_i$ denies kidney from $D_j$ and turns down the offer.

In order to solve this problem, we ask for acceptable communities from patients in the initial stages of IKEP. 

\subsection{Societal Acceptance Score($sas_{ij}$)} 
\label{sasSubsection}

It is the measurement of the societal acceptance of donor's kidney by the patient. It is calculated by the \cref{eq:saScore}.

\begin{equation}
    \label{eq:saScore}
    sas_{ij} = 
    \begin{dcases}
        1,                 & \text{if } sdPref_j=\emptyset\\
        \frac{1}{k},       & \text{if } sd_i\in sdPref_j\\
        \frac{1}{k^*},      & \text{otherwise}
    \end{dcases}
\end{equation}

Where, $k$ is index of donor's societal distribution in patient's preference list. And $k^*$ is any number which is greater than $n_{sd}$.

The \cref{algo:saScore} implements SAS calculation. It checks the length of preference list. If found 0, it returns 1 or else, it traverses through the preference list of the patient. The element at index $k$ of the preference list which matches with the societal distribution of donor is returned as $\frac{1}{k}$. In case the donor's societal distribution doesn't exist in the preference list\footnote{It is assumed that, if any patient has a set of preferences, $\forall sd\in SD-sdPref_j$ is unacceptable for that particular patient.}, the value returned $\frac{1}{k^*}$.

\begin{algorithm}[H]
  \caption{Calculate Societal Acceptance Score}\label{algo:saScore}
  \begin{algorithmic}[1]
    \Procedure{sas}{$sd_i,sdPref_j$}\Comment{Calculate societal acceptance score}
    \State \textbf{Initialize} $k = 1$
    \If{$sdPref_j.$length()$=0$}
        \State \Return{1}
    \Else
        \ForAll{$s \in sdPref_j$}
            \If{$s=sd_i'$}
                \State \Return{$\frac{1}{k}$}
            \EndIf
            \State $k = k+1$
        \EndFor
        \State \Return{$\frac{1}{k^*}$}
    \EndIf
    \EndProcedure
  \end{algorithmic}
\end{algorithm}

The value returned by the \cref{algo:saScore} is multiplied with the general perspective weight calculated in \cref{eq:generalWeight}. 

\subsection{Exclusion Coefficient($ec_i$)}
\label{exclusionCoefficient}
Here we propose a mechanism for checking the foul plays and stopping fraudulent players form manipulating outcomes in the market by misrepresenting their $sdPref$(s). 
\subsubsection{Why Exclusion Coefficient?}
Pairs might choose to reject the kidneys based on their societal ideologies even after choosing, not to show them up in $sdPref$ in the first place, in order to improve their chances of getting selected in exchange. 

\begin{figure}[H]
\centering
\includegraphics{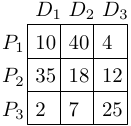}
\caption{Compatibility Matrix Before SAS}
\label{fig:compatibilityMatrixEC_beforesas}
\end{figure}

Let there be 3 pairs in a pool, $(P_1,D_1)$, $(P_2,D_2)$ and $(P_3,D_3)$ and let \cref{fig:compatibilityMatrixEC_beforesas} be the compatibility matrix just before the implementation of $sas$ i.e, $w_{ij}= gw_{ij}$.

All the $gw_{ij}$ are arbitrarily taken, in order to depict a realistic situation and eventually visualize the significance of $ec_i$.

Let there be 2 societal distributions, $sd_1$, $sd_2$. Let $(P_1,D_1)$ belong to societal distribution $sd_2$ and $(P_2,D_2),(P_3,D_3)$ belong to societal distribution $sd_1$\footnote{\label{whysamesd}It is realistic, generally, that $sd_{P_i}=sd_{D_i}$ as in Indian context, donor is a close relative of patient. We don't take inter societal distributional relatives into consideration for easy visualization.Although, in the case of inter societal distributional relatives, the societal distribution of patient is irrelevant, since, he/she is the one who is accepting the kidney. Societal distribution of donor is important though.}. And let $F^*=19$. Let there be 2 situations,

\begin{enumerate}
    \item $(P_2,D_2)$ represents $sdPref$ truthfully: 
    
    Let the respective $sdPref$(s) for $(P_1,D_1)$, $(P_2,D_2)$ and $(P_3,D_3)$ be $\{sd_2,sd_1\},$ $\{sd_1,sd_2\}$ and $\{sd_1,sd_2\}$. 
        
    \begin{figure}[H]
    \centering
    \includegraphics{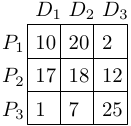}
    \caption{Compatibility Matrix After SAS}
    \label{fig:compatibilityMatrixEC_aftersas}
    \end{figure}
    
    \cref{fig:compatibilityMatrixEC_aftersas} represents compatibility matrix after the inclusion of sas, when $(P_2,D_2)$ represents his/her $sdPref$ truthfully, using $w_{ij}= gw_{ij} \times sas_{ij}$\footnote{The calculations regarding the generation of $sas_{ij}$ has be discussed in \cref{sasSubsection} and the equation govering the score is stated in \cref{eq:saScore}}.
    
    \cref{fig:compatibilityGraphEC_afterSAS} is the compatibility graph representing the top most preferences from \cref{fig:compatibilityMatrixEC_aftersas} and the possible exchange cycles present. $(P_3,D_3)$ gets into a self loop and gets out of the pool, whereas $(P_1,D_1)$ and $(P_2,D_2)$ remain in the pool for waiting for upcoming new participants.
    
    \begin{figure}[H]
    \centering
    \includegraphics[width=8cm]{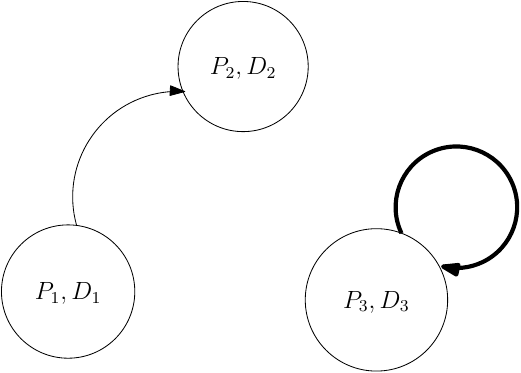}
    \caption{Compatibility Graph After SAS}
    \label{fig:compatibilityGraphEC_afterSAS}
    \end{figure}
    
    \item \textbf{$(P_2,D_2)$ misrepresents $sdPref$}. Let the respective $sdPref$(s) for $(P_1,D_1),(P_2,D_2)$ and $(P_3,D_3)$ be $\{sd_2,sd_1\},\emptyset$ and $\{sd_1,sd_2\}$, where $(P_2,D_2)$ misrepresents. And, let the true $sdPref_{(P_2,D_2)}$ be, as same as what specified in situation 1.
    \begin{figure}[H]
    \centering
    \includegraphics{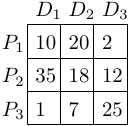}
    \caption{\centering Compatibility Matrix After SAS When $(P_2,D_2)$ misrepresents}
    \label{fig:compatibilityMatrixEC_aftersas_misrepresenting}
    \end{figure}
    
    \cref{fig:compatibilityMatrixEC_aftersas_misrepresenting} represents the compatibility matrix after calculating modified $w$(s) using the $sdPref$(s) mentioned.
    
    \begin{figure}[H]
    \centering
    \includegraphics[width=8cm]{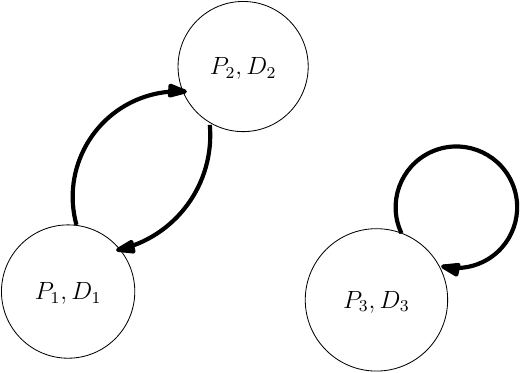}
    \caption{\centering  Compatibility Graph After SAS When $(P_2,D_2)$ misrepresents.}
    \label{fig:compatibilityGraphEC_afterSAS_misrepresents}
    \end{figure}
    
    \cref{fig:compatibilityGraphEC_afterSAS_misrepresents} is the compatibility graph representing the top most preferences and proposed exchanges from \cref{fig:compatibilityMatrixEC_aftersas_misrepresenting}. The proposed cycles would be $(P_1,D_1)\longrightarrow(P_2,D_2)\longrightarrow(P_1,D_1)$ and $(P_3,D_3)\longrightarrow(P_3,D_3)$.
\end{enumerate}

$(P_2,D_2)$ after misrepresenting its $sdPref$, gets included in the cycles, unlike what happens when $(P_2,D_2)$ truthfully states $sdPref$ in the first situation, which motivates $(P_2,D_2)$ to falsely state $sdPref= \emptyset$ with the idea of initially getting into a cycle and then either rethinking the exchange to be emotionally, culturally or societally feasible. 

Families might get upset about the decision taken by the pair. Even the donor can choose to quit, over the disagreement over the societal distributions, which in turn might force the patient to reject the proposed kidney. Hence, it is better for the patients to get in terms with their donors before putting up their $sdPref$(s).

\subsubsection{How does it works?}
This strategy is countered by excluding those pairs in the upcoming IKEPA runs, introducing a new value, named as Exclusion Coefficient(ec)$\in [0,1]$.

The main idea to calculate the value $ec_i$ is, to reduce the willingness of all the pairs $j\neq i$, from accepting the kidney, from the pair $i$.

\begin{equation}
    \label{eq:ecScore}
    ec_i = 
    \begin{dcases}
        \frac{1}{k_{ec}},   & \text{if } rar = 1\\
        \frac{1}{k_{ec}-R^{'}\times (rar-1)},   & \text{if } rar \leq R_{ec}\\
        1,      & \text{otherwise}
    \end{dcases}
\end{equation}

\begin{equation}
    \label{eq:reducingCoefficient}
    R^{'}= \frac{(k_{ec}-1)}{R_{ec}}
\end{equation}

Here, $rar$ is the IKEPA runs executed, after rejecting the last kidney exchange proposed, by the pair $j$. $k_{ec}$ the constant value by which the final weight needs to be divided in order to exclude the pair from further $R_{ec}$ runs. $R^{'}$ is the reducing coefficient calculated in \cref{eq:reducingCoefficient}. And in order to slighty reduce the exclusion by each IKEPA run, we subtract $k_{ec}$ by  $R^{'}$ times the number of runs performed after rejections. We take $k_{ec}$ as 10, and $R_{ec}$ as 4.

Hence, the final \textbf{weight} of edge $e_{ij}$ is calculated as:

\begin{equation}
    \label{eq:finalWeight}
    w_{ij} = gw_{ij} \times sas_{ij} \times ec_{i}
\end{equation}

Here, $i$ as specified earlier, denotes pair which is providing the kidney, and $j$ denotes the pair receiving the kidney. The value of $ec_i$ is calculate before each IKEPA run after all the rejection data is received.

For example, let a patient $P_1$ has $sdPref_1 = [sd_3,sd_1]$ and a donor $D_3$ belonging to a societal distribution $sd_1$. Let the general perspective weight $w_{31}=30$. Since the index of societal distribution of $D_3$ in the societal preference of patient $sdPref_1$ is 2, the $sas_{31}= \frac{1}{2}$. Now, let us assume that the pair 3 has rejected an offer in the last IKEPA run, hence the excluding coefficient $ec_3= \frac{1}{10}$ in the first run after rejection. And, the final weight $w_{31} = 30 \times \frac{1}{2} \times \frac{1}{10}$, i.e, 1.5.

In the upcoming IKEPA runs, the values of $ec_3$ will be $\frac{1}{7.75}$, $\frac{1}{5.5}$, $\frac{1}{3.25}$ and $1$ respectively. And, the final weight $w_{31}$ would look like, 1.94, 2.73, 4.62 and 15 respectively in each upcoming IKEPA run.

\subsection{Generation Of Compatibility Matrix: Extended}
The \cref{algo:genCompatibilityMatrix_extended} shows, how the final weight is been calculated. The \cref{algo:genCompatibilityMatrix_extended} needs to replace \cref{algo:genCompatibilityMatrix} stated in \cref{IKEP_algorithms}. 

IKEP at the time of edge selection in \cref{algo:ikep_genGraphMaxRowVal}, checks $w_{ij}$ to be greater than $F^*$. Here $F^*$ is the filtering constant to remove \textbf{"not up to the mark"} edges. The not up to the mark value is dependent upon the thickness of the pool. In situations with scarce participants, $F^*$ will go down and vice-versa for a surplus in participants.

\begin{algorithm}[H]
  \caption{Extended Generate Compatibility Matrix}\label{algo:genCompatibilityMatrix_extended}
  \begin{algorithmic}[1]
    \Require{$V_{1} \dots V_{N}$} 
    \Procedure{genCompatibilityMatrix}{$V$}\Comment{Generate Compatibility Graph}
    \State \textbf{Initialize} $W(e_{ij}) = 0$, for all $e_{ij} \in E$
    \ForAll{$i\in V$}
    \ForAll{$j\in V$}
    \State $b_{ij} = aboScore(DB_i,B_j)$
    \If{$b_{ij}=0$}
        \State $w_{ij} = 0$
        \State \textbf{continue}
    \EndIf
    \State $h_{ij} = hlaScore(DH_i,H_j)$
    \If{$h_{ij}=0$}
        \State $w_{ij} = 0$
        \State \textbf{continue}
    \EndIf
    \State $a_{ij} = ageScore(DAge_i,Age_j)$
    \State $k_{ij} = kidneyScore(DK_i,K_j)$
    \State $p_{ij} = pinScore(DPin_i,Pin_j)$
    \State $gw_{ij} = h_{ij}+b_{ij}+k_{ij}+a_{ij}+p_{ij}$
    \State $sas_{ij} = sas(Dsd_i,sdPref_j)$
    \State $w_{ij} = gw_{ij} \times sas_{ij} \times ec_i$
    \EndFor
    \EndFor
    \State \Return{$W$}
    \EndProcedure
  \end{algorithmic}
\end{algorithm}

\begin{figure}[H]
\centering
\includegraphics{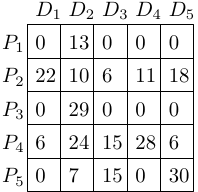}
\caption{Final Compatibility Matrix}
\label{IKEA_compatibilityMatrixWithIndianParameter}
\end{figure}

Let there be 3 societal distributions prevailing in India. 
\[SD_{Ind} = \{sd_1,sd_2, \dots sd_{3}\}\]
Let the \cref{tab:exampleCharacteristicTable_extended} be the extended version of \cref{tab:exampleCharacteristicTable} with added patients' societal preferences and societal distribution of donors. \cref{IKEA_compatibilityMatrixWithIndianParameter} is the modified compatibility matrix of \cref{IKEA_compatibilityGraph} with final weights calculated by \cref{eq:finalWeight}.

\begin{sidewaystable}
\caption{\cref{tab:exampleCharacteristicTable} with societal preferences}
\label{tab:exampleCharacteristicTable_extended}
\vspace{2 mm}
\centering
\begin{tabular}{c c c c c c c c c} \hline \\
Name  & Blood & HLA &Age &Kidney &PinCode &Societal Pref.(for patient) &Init. &Top \\ 
&Group&&& Size&&/Societal Dist.(for donor) &Priority&Pref.\\ \hline
$P_1$ & $A^-$  & A1,B8,DR10,A3,B14,DR17 &45 &11 &496001 &$[sd_1,sd_2]$ &\multirow{2}{*}{$5$} &\multirow{2}{*}{$D_2$}\\
$D_1$ & $B^+$  & A2,B7,DR11,A10,B16,DR8 &30 &12 &496001 &$sd_1$\\
$P_2$ & $AB^+$ & A1,B8,DR10,A2,B7,DR11 &25 &10 &490020 &$[sd_1,sd_2]$ &\multirow{2}{*}{$2$} &\multirow{2}{*}{$D_1$}\\
$D_2$ & $O^-$  & A1,B8,DR10,A10,B16,DR8 &55 &11.5 &496001 &$sd_2$\\
$P_3$ & $A^-$  & A1,B8,DR17,A10,B16,DR8 &67 &12 &496001 &$\emptyset$ & \multirow{2}{*}{$1$} &\multirow{2}{*}{$D_2$} \\
$D_3$ & $A^+$  & A1,B8,DR17,A10,B16,DR8 &25 &11 &496001 & $sd_3$\\
$P_4$ & $AB^+$ & A1,B8,DR17,A10,B16,DR8 &30 &10 &496001 &$[sd_2,sd_3]$ & \multirow{2}{*}{$6$} &\multirow{2}{*}{$D_4$} \\
$D_4$ & $B^-$  & A1,B8,DR17,A10,B16,DR8 &27 &12 &496001 & $sd_2$\\
$P_5$ & $A^+$  & A1,B8,DR17,A10,B16,DR8 &60 &11.5 &496001 &$[sd_1,sd_3]$ & \multirow{2}{*}{$10$} &\multirow{2}{*}{$D_5$}\\
$D_5$ & $O^+$  & A1,B8,DR17,A10,B16,DR8 &55 &11.5 &496001 &$sd_1$\\ \hline
\end{tabular}
\end{sidewaystable}

\begin{figure}[H]
\centering
\includegraphics[width=8cm]{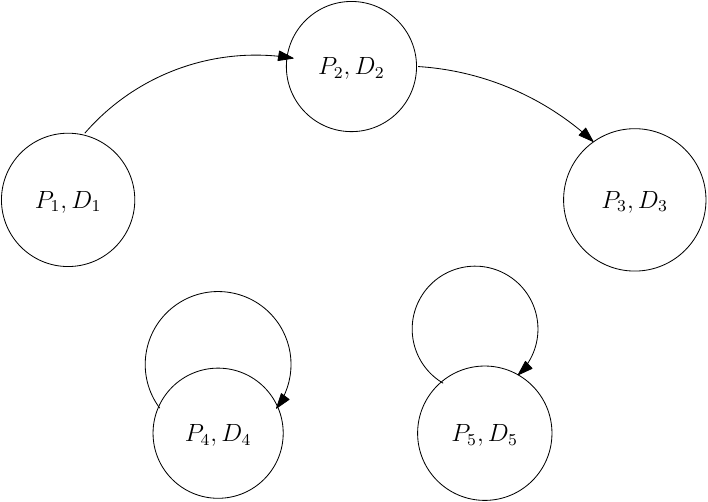}
\caption{Final Compatibility Graph}
\label{IKEA_compatibilityGraphWithIndianParameter}
\end{figure}

\cref{IKEA_compatibilityGraphWithIndianParameter} is the graph representation of \cref{IKEA_compatibilityMatrixWithIndianParameter}\footnote{Every edge from vertex $v_1$ to $v_2$ signify that $v_2$'s top preferred kidney is from pair $v_1$.}. Where $(P_4,D_4)$ and $(P_5,D_5)$ have self loops, preferring their own kidneys only. Which in contrast to earlier preferences as shown in \cref{IKEA_compatibilityGraph}, where they preferred $D_3$, the most.

\begin{figure}[H]
\centering
\includegraphics[width=8cm]{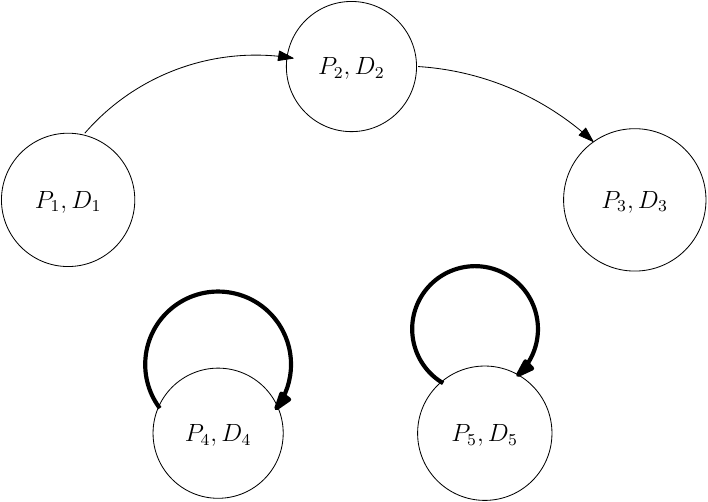}
\caption{\centering Proposed Cycles in First \& Second Iteration}
\label{IKEA_compatibilityGraphWithIndianParameter_FirstIteration}
\end{figure}

\cref{IKEA_compatibilityGraphWithIndianParameter_FirstIteration} proposes two exchanges. In the first two iteration the pairs $(P_4,D_4)$ and $(P_5,D_5)$ are proposed as cycles and removed. And unfortunately, the pair $(P_1,D_1)$, $(P_2,D_2)$ and $(P_3,D_3)$ will need to wait for the next time untill another patient-donor pair enroll in to the IKEP. Which contradicts to the results shown by \cref{IKEA_compatibilityGraph}, since $(P_2,D_2)$ strongly disgraces the societal distribution $sd_3$ of $(P_3,D_3)$.

\begin{figure}[H]
\centering
\includegraphics{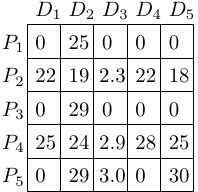}
\caption{\centering Compatibility Matrix After Cycle Rejection}
\label{IKEA_compatibilityMatrix_exclusionCoefficient}
\end{figure}

Now, in order to understand the effect of $ec$, let us reconsider the example stated in \cref{IKEP_algorithms}. Let the pair $(P_3,D_3)$, turn down the exchange, proposed in \cref{IKEA_compatibilityGraphProposedCycle}. Hence, $ec_3= 0.1$ the compatibility matrix for next IKEPA run would look like as shown in \cref{IKEA_compatibilityMatrix_exclusionCoefficient}.

\begin{figure}[H]
\centering
\includegraphics[width=8cm]{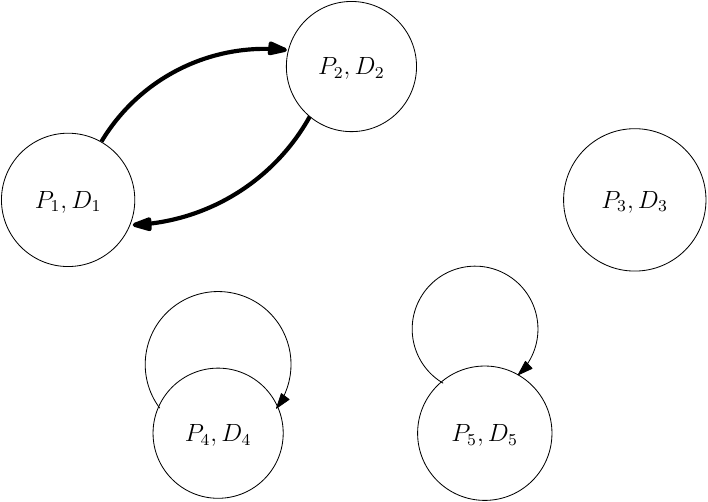}
\caption{\centering Proposed Cycles In Second Iteration After Rejection By $(P_3,D_3)$}
\label{IKEA_compatibilityGraph_exclusionCoefficient_FirstIteration}
\end{figure}

According to \cref{IKEA_compatibilityMatrix_exclusionCoefficient}, the new compatibility graph is shown in \cref{IKEA_compatibilityGraph_exclusionCoefficient_FirstIteration}, which proposes a new cycle $(P_1,D_1)\longrightarrow (P_2,D_2)\longrightarrow (P_1,D_1)$ and rest remains the same. As a result of which, $(P_3,D_3)$ looses their chance of getting matched and hence, will have to wait for upcoming IKEP runs, when other pairs come in.

\begin{figure}[H]
\centering
\includegraphics{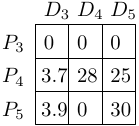}
\caption{\centering Compatibility Matrix For Third Iteration}
\label{IKEA_compatibilityMatrix_exclusionCoefficient2}
\end{figure}

\begin{figure}[H]
\centering
\includegraphics[width=8cm]{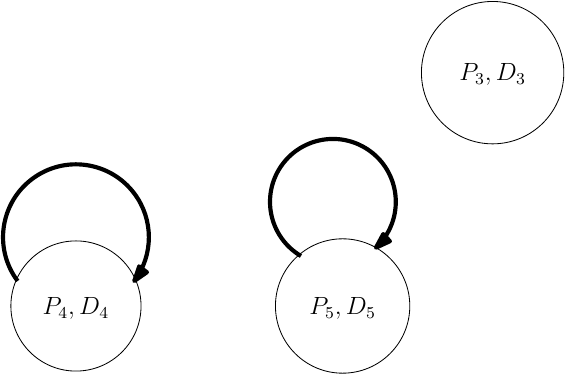}
\caption{\centering Proposed Cycles in Third Iteration}
\label{IKEA_compatibilityGraph_exclusionCoefficient_SecondIteration}
\end{figure}

After removal of participating vertices in the proposed cycles, \cref{IKEA_compatibilityMatrix_exclusionCoefficient2} is generated as the compatibility matrix for next IKEPA iteration with $ec_3= \frac{1}{7.75}$. \cref{IKEA_compatibilityGraph_exclusionCoefficient_SecondIteration} shows the next two cycle proposals in the upcoming iterations, namely, $(P_4,D_4)$ and $(P_5,D_5)$. And at last, unlike previous match in which $(P_1,D_1)$ was left unmatched, $(P_3,D_3)$ is left unmatched with $ec$ in action.

Now we discuss one of the shortcomings of IKEP and the disadvantages of ways of usage of $F^*$. And then propose an enhanced version of IKEPA.

\section{Effects of Edge Filter and Enhanced IKEPA}
\label{effectsEFAndEIKEPA}

\subsection{How Edge Filter works?}
\begin{figure}[H]
\centering
\includegraphics[width=3.5cm]{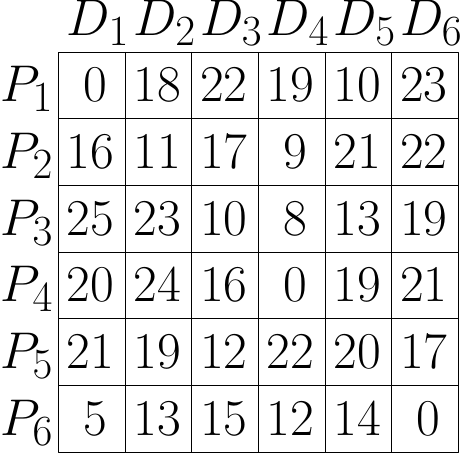}
\caption{\centering Example Compatibility Matrix}
\label{IKEA_compatibilityMatrix_effectsofedgefilter}
\end{figure}

\begin{figure}[H]
\centering
\includegraphics[width=5cm]{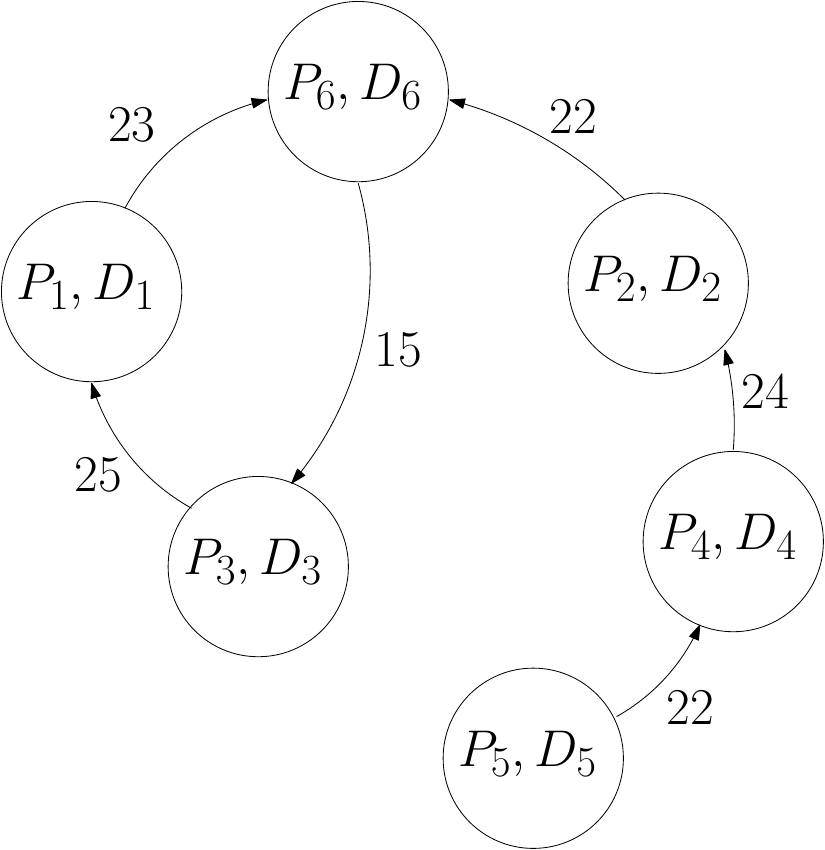}
\caption{\centering Compatibility Graph without the influence of Edge Filter}
\label{IKEA_compatibilityGraph_effectsofedgefilter}
\end{figure}

Let us take an exemplar compatibility matrix as described in \cref{IKEA_compatibilityMatrix_effectsofedgefilter}. And let \cref{IKEA_compatibilityGraph_effectsofedgefilter} be the graphical representation of \cref{IKEA_compatibilityMatrix_effectsofedgefilter}.

\begin{figure}[H]
\centering
\includegraphics[width=5cm]{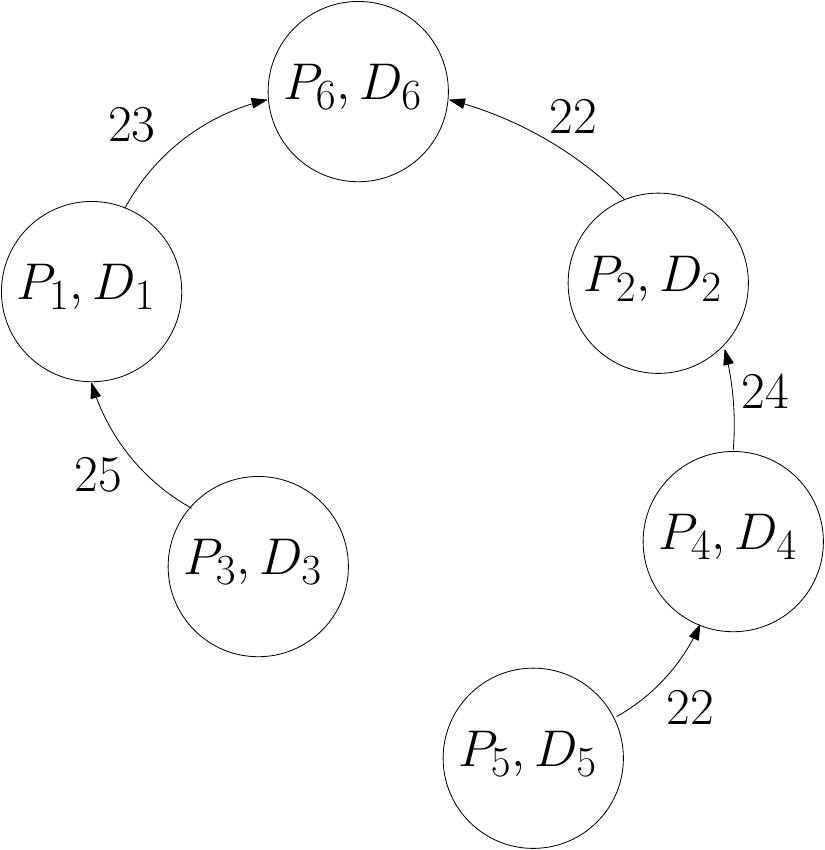}
\caption{\centering Compatibility Graph with the influence of Edge Filter in IKEPA}
\label{IKEA_compatibilityGraph_effectsofedgefilter_effect}
\end{figure}

According to \cref{IKEA_compatibilityGraph_effectsofedgefilter}, there is only one possible cycle, that can be figured out from the graph, i.e, $(P_1,D_1)$ $\longrightarrow$ $(P_6,D_6)$ $\longrightarrow$ $(P_3,D_3)$ $\longrightarrow$ $(P_1,D_1)$. But this cycle gets rejected by \cref{algo:ikep_genGraphMaxRowVal} Line 5\footnote{By not adding an adjacent node i.e, $(P_3,D_3)$ here, to the vertex $(P_6,D_6)$.}. \cref{IKEA_compatibilityGraph_effectsofedgefilter_effect} shows the final graph created by \cref{algo:ikep_genGraphMaxRowVal} when $F^*=20$. As a result of which all the potential pairs are not able to match\footnote{Due to the incompetency of $(P_6,D_6)$, having below the $F^*$, top preferred patient in the pool}. And since, none of the pairs are getting matched, this cycle remains as it is, till the end of IKEP and can only change, when any other pair enter in to the market and influences the cycle in such a way that new, non-rejectable cycles are formed.

The weights on the edges between two nodes are the compatibility score of the kidney from the donor of the node at the head of the arrow, with the patient of the node at the tail of the arrow.

\subsection{E-IKEPA}
\label{enhancedIKEPA}
In order to stop all the potential pairs from falling out of the cycles, because of the cycles getting rejected, because of the below the threshold edges, we propose an enhanced form of IKEPA, named as E-IKEPA. The main idea behind E-IKEPA is a pre-processed top trading cycle, i.e, as shown in \cref{algo:CompatibilityMatrixFilter}, all the nodes, with maximum compatibility score lesser than $F^*$ are removed out of the pool. And after that all the cycles are formed by the remaining pairs in the pool.

\begin{algorithm}[H]
  \caption{E-IKEPA}\label{algo:ikep_main_enhanced}
  \begin{algorithmic}[1]
    \Procedure{E-IKEPA}{$W,S,V$}\Comment{Removes all those pairs, whose top preferred nodes, do not qualify edge filter threshold}
        \State \textbf{initialize} $C= new\ ArrayList<>()$
      \While{$V \neq null$}\Comment{End, if no vertices left}
      
        \State $W = $GenCompatibilityMatrix$(V)$
        \State $W,V=$FilteredData$(W,V)$
        \State $Q = $ArrayToQueue$(V)$
        \State $c = $FindCycle$(W,V.size,Q)$
        \If{$c \neq null$}
            \State RemoveCycle$(C,V,W)$
            \State $C.$add$(c)$
        \Else 
            \State $Break$
        \EndIf
      \EndWhile\label{ikepWhile_Exhanced}
      \State \textbf{return} $C$\Comment{List of cycles}
    \EndProcedure
  \end{algorithmic}
\end{algorithm}

\cref{algo:CompatibilityMatrixFilter} filters the inadequate nodes from the market, using an infinite while loop, which removes all those nodal rows which have max compatibility score lesser than the $F^*$ and also the node specific entries in the remaining rows. And, after that, those vertices are also removed from the list of vertices.

\begin{algorithm}[H]
  \caption{FilterData}\label{algo:CompatibilityMatrixFilter}
  \begin{algorithmic}[1]
    \Procedure{FilteredData}{$W,V$}\Comment{Generates new matrix with quality assured pairs}
      \While{TRUE}
        \State $W'=[][]$
        \ForAll{$w_i \in W$}\Comment{$w_i$ is $i^{th}$ row in adjacency matrix}
        \State $i,j = $max$(w)$ \Comment{if max is non-distinct, use pair priority to break ties.}
        \If{max$(w)< F^*$}
        \State $W.$removeRow$(w)$
        \State $W.$removeAllEntriesWith$(w)$
        \State $V$.remove(V[i])
        \EndIf
        \EndFor
        \If{$isOptimized(W',V.size())$}
        \State \textbf{return} $W',V$
        \Else
        \State $W=W'$
        \EndIf
      \EndWhile
      \State \textbf{return} $C$\Comment{List of cycles}
    \EndProcedure
  \end{algorithmic}
\end{algorithm}
\cref{algo:isOptimized} iterates through each and every value of the compatibility matrix, and checks if any of the max value of the rows do not fulfil the $<F^*$ condition. When found any, \textbf{FALSE} is returned.
\begin{algorithm}[H]
  \caption{isOptimized}\label{algo:isOptimized}
  \begin{algorithmic}[1]
    \Procedure{isOptimized}{$W,V$}\Comment{Checks if the compatibility matrix is optimized}
      \ForAll{$w_i \in W$}\Comment{$w_i$ is $i^{th}$ row in adjacency matrix}
        \State $i,j = $max$(w)$ \Comment{if max is non-distinct, use pair priority to break ties.}
        \If{max$(w)< F^*$}
        \State \textbf{return} FALSE
        \EndIf
        \EndFor
        \State \textbf{return} TRUE
    \EndProcedure
  \end{algorithmic}
\end{algorithm}

\begin{figure}[H]
\centering
\includegraphics[width=3cm]{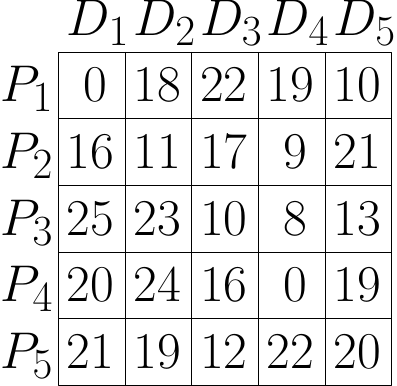}
\caption{\centering Enhanced Compatibility Matrix}
\label{IKEA_compatibilityMatrix_enhanced}
\end{figure}

\begin{figure}[H]
\centering
\includegraphics[width=6cm]{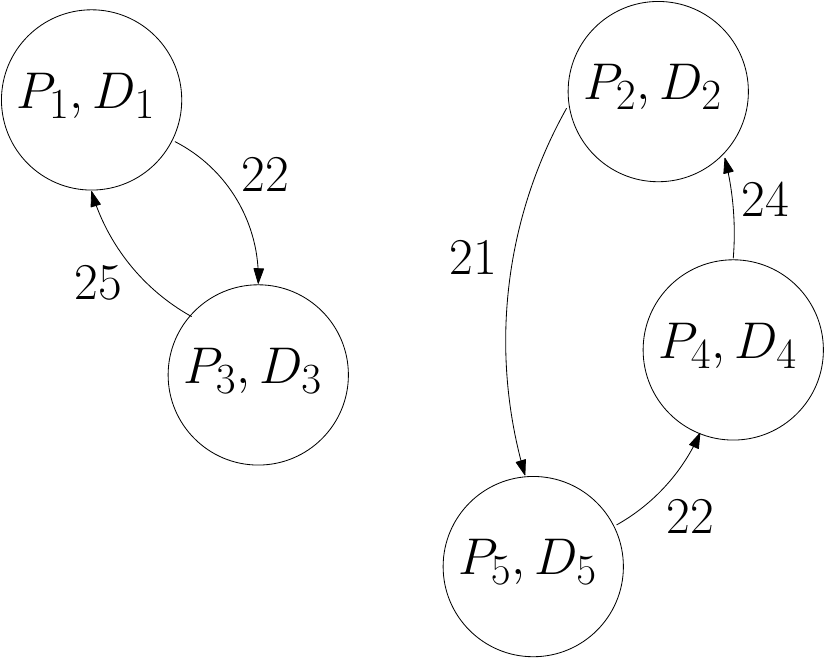}
\caption{\centering Enhanced compatibility graph}
\label{IKEA_compatibilityGraph_enhanced}
\end{figure}

Hence, according to \cref{algo:ikep_main_enhanced}, the node $(P_6,D_6)$ is removed out for this IKEP run and the enhanced compatibility matrix is shown in \cref{IKEA_compatibilityMatrix_enhanced} and \cref{IKEA_compatibilityGraph_enhanced} shows the graphical representation of the same.

\cref{IKEA_compatibilityGraph_enhanced} now proposes two cycles, i.e, $(P_1,D_1)$ $\longrightarrow$ $(P_3,D_3)$ $\longrightarrow$ $(P_1,D_1)$ and $(P_2,D_2)$ $\longrightarrow$ $(P_5,D_5)$ $\longrightarrow$ $(P_4,D_4)$ $\longrightarrow$ $(P_2,D_2)$, matching all the pairs, except $(P_6,D_6)$.

Hence, E-IKEPA improves $pair^m$ by a considerable amount and the acceptance of exchanges proposed. \cref{eikepaSimulationResults} shows the increase in $pair^m$ by E-IKEPA.

\section{Breaking ties through ranks}
\label{prioritizingPairs}
In this section, we discuss how the preferences stated by the participants are made strict\footnote{by breaking the ties between the preference caused by same compatibility score.}

Since IKEP is a variant of Gale's top trading cycle algorithm, priority doesn't plays any significant role. The only place where priority seems to benefit, is selection of cycles. Since, TTCA is based on strict preferences($\succ$), no vertex could have more than one outgoing edge\footnote{ Outgoing edge's ending vertex shows the most preferred donor/pair for the starting vertex of that particular outgoing edge.}. As a result of this, no vertex could participate in more than one cycle, which saves them from having conflicts in getting selected in a particular cycle.

The main requirement of prioritizing pairs is to break the ties and make the patient's preference on the donors, strict preferences. The pairs are prioritized based on the how difficult is for them to get a compatible donor. There priorities are used to break ties between the pairs who appear to be identical to the patient. The priority is calculated based on the following factors.

\subsection{Generic factors}
The generic factors specified and their scoring system has been inspired from the Updated Kidney Allocation Policy\cite{39}
\subsubsection{PRA score.} Higher the PRA score of the patient, higher is the priority given to the pair.
\begin{equation}
    \label{eq:praPriority}
    PRA = 
    \begin{dcases}
        0,              & \text{if } PRA_j\leq D_{pra}\\
        (PRA_j-D_{pra}) \times \alpha_{pra},    & \text{otherwise}
    \end{dcases}
\end{equation}

Where $D_{pra}$ is any PRA value below which, if found is considered insignificant. Hence the PRA score is set to zero. Or else, The $D_{pra}$ is subtracted from the PRA of patient and multiplied by a constant $\alpha_{pra}\in [0,1]$. We take value of $D_{pra}$ as 20 and the value of $\alpha_{pra}$ as 0.05.

\subsubsection{Level of Difficulty to match pairs} is used to determine, how much priority should these pairs be endowed with. Pair types O-A, O-B, O-AB, A-AB and B-AB do offer less demanded kidneys and seek highly demanded kidneys, hence wait for longer periods. Since, they have a disadvantage, should be prioritized marginally.
\begin{equation}
    \label{eq:pairTypePriority}
    PT = 
    \begin{dcases}
        V_{pt},    & \text{if } Pair\ Type\in \{O-A,O-B,\\
                            & O-AB,A-AB,B-AB\}\\
        0,    & \text{otherwise}
    \end{dcases}
\end{equation}

Where $V_{pt}\in (0,\infty)$ be any number, which adds up to the priority score in order to provide an advantage to the patients with lower chances of getting matched in IKEP. We take the value of $V_{pt}$ as 1.

\subsubsection{Age.} The children and young adults are given more priority, since they are supposed to get more benefits than the older patients. For the older participants, including above stated factors, negative outcomes from previous algorithm runs and the patient who was previously a donor are some of the significant factors for prioritization.
\begin{equation}
    \label{eq:agePriority}
    Age = 
    \begin{dcases}
        V_{ap}-\alpha_{ap}\times D_{ap},  & \text{if } Age_j\in List_{Age}\\
        0,    & \text{otherwise}
    \end{dcases}
\end{equation}
Where $V_{ap}$ is the upper limit of Age priority score. $List_{Age}$ is the list of Age groups. $D_{ap}$ be the index of Age group in which $Age_j$ lies in. 

$\alpha_{ap}$ is the regulator of the decrease in the value of Age priority score with the increase in the index of the age group in which $Age_j$ lies in, such that.

\[\alpha_{ap}\in (0,\frac{V_{ap}}{List_{Age}.length()})\]

We assume the value of $V_{ap}$ as 3 and $\alpha_{ap}$ as 1. The list of age groups taken is $List_{Age} = [<6,$ $<12,$ $<18]$.

\subsubsection{Waiting Time.} Time is expensive, hence need to be used to prioritize patients. Higher the waiting time of patient, higher is the priority. For every algorithm run with negative result, there will be an increase in priority by $V_{wt}$ units. The initial waiting time priority for new participants will be zero. Although, a provision should be made for those who have been in the waiting list for deceased kidney in order to unify the two models. We take value $V_{wt}$ as 1. 
\subsubsection{Is undergoing temporary vascular access}
There are two situations prevailing among the patients already undergoing hemodialysis.

\textbf{Situation 1:} One, who has failed on all AV Fistula sites.

\textbf{Situation 2:} One, who has failed AV Graft after failing all AV Fistula sites.

\begin{equation}
    \label{eq:vaPriority}
    VA = 
    \begin{dcases}
        V_{vap},    & \text{if } Situation\ 2\ applies\\
        V_{vap1},    & \text{if } Situation\ 1\ applies\\
        0,    & \text{otherwise}
    \end{dcases}
\end{equation}

Where, $V_{vap}$, $V_{vap1}$ be the scores assigned to VA priority score according to the conditions they fulfil such that, $V_{vap}\geq V_{vap1}$. We assume the values of $V_{vap}$ and $V_{vap1}$ as 6 and 2 respectively.

\subsubsection{Patient who was previously a donor} One who had been a donor and has now lost the second kidney needs to be prioritized.
\begin{equation}
    \label{eq:ipdPriority}
    IPD = 
    \begin{dcases}
        V_{ipd},     & \text{if } Patient\ was\ once\ a\ donor\\
        0,    & \text{otherwise}
    \end{dcases}
\end{equation}

Where $V_{ipd}\in (0,\infty)$ is the maximum score allocated to the IPD priority. We take $V_{ipd}$ as 5.

\subsection{IKEP specific factors:} The above stated factors are commonly used in kidney exchange/deceased kidney waiting list priorities for the participants. There are some additional factors, which will influence the scoring system for prioritizing the patients in IKEP.
\subsubsection{Distance from dialysis center.} In India, a huge amount\footnote{According to the \href{https://censusindia.gov.in/census_data_2001/india_at_glance/rural.aspx}{\emph{census in 2001}}, 72\% of the Indian population resided in urban areas.} of population resides in rural distant villages with insignificant health care systems and tedious transport system.
    
    \begin{equation}
        \label{eq:distPriority}
        Dist = 
        \begin{dcases}
            V_{d}-\alpha_{d}\times D_{d},  & \text{if } Dist_j \in List_{Dist}\\
            0,    & \text{otherwise}
        \end{dcases}
    \end{equation}
    
    Where $V_{d}$ is the upper limit of distance priority score. $List_{Dist}$ is the list of distance ranges. $D_{d}$ be the index of distance group in which $Dist_j$ lies in.

    $\alpha_{d}$ is the regulator of the decrease in the value of distance priority score with the increase in the index of the distance group in which $Dist_j$ lies in, such that.
    
    \[\alpha_{d}\in (0,\frac{V_{d}}{List_{Dist}.length()})\]
    
    We assume the value of $V_{d}$ as 3 and $\alpha_{d}$ as 1. The list of age groups taken is $List_{Dist} = [\geq 50,$ $>10\And<50]$.
    
    All the values presented are in kilometers.
    
\subsubsection{Economic slab of the patient.} There is also an increasing economic gap in India. Rural people with limited income and lesser knowledge and availability of insurance and other aids, find it difficult to stay on dialysis. And hence need to be prioritized \emph{marginally}.
    
    \begin{equation}
        \label{eq:economicPriority}
        Eco = 
        \begin{dcases}
            V_{eco}-\alpha_{eco}\times D_{eco},  & \text{if } Eco_j \in List_{eco}\\
            0,    & \text{otherwise}
        \end{dcases}
    \end{equation}
    
    Where $V_{eco}$ is the upper limit of economic priority score. $List_{eco}$ is the list of economic slab. $D_{eco}$ be the index of economic slab in which $Eco_j$ lies in.

    $\alpha_{eco}$ is the regulator of the decrease in the value of economic priority score with the increase in the index of the economic slab in which $Eco_j$ lies in, such that.
    
    \[\alpha_{eco}\in (0,\frac{V_{eco}}{List_{eco}.length()})\]
    
    We assume the value of $V_{eco}$ as 4 and $\alpha_{eco}$ as 1. The list of age groups taken is 
    
    $List_{Eco} = [<1,$ $>1\And<5,$ $>5\And<10,$ $\geq 10]$.
    
    All the values presented are in Lakhs.

\subsubsection{Priority Allocation} The initial priority $Priority_v$ of the pair is calculated with the help of the following formula. And for the subsequent calculations, we need to increment the value of Waiting time score(WTScore). 

\begin{equation}
    \label{eq:initPriority}
    priority_{ij} = Age+PRA+PT+VA+IPD+Dist+Eco
\end{equation}

\begin{algorithm}[H]
  \caption{Assigns Priority}\label{algo:assignInitialPriority}
  \begin{algorithmic}[1]
    \Require{$V_{1} \dots V_{N}$}
    \Procedure{calculatePriority}{$V$}\Comment{Assigns Priority}
    \ForAll{$v \in V$}
        \If{$isInitial_v$}
            \State $WTScore_v = 0$
            \State $isInitial_v = false$
        \Else
            \State $WTScore_v = WTScore_v+V_{wt}$
        \EndIf
        \State  $Priority_v = Age+PRA+PT+VA+IPD+Dist+Eco+WTScore_v$
    \EndFor
    \State \Return{$V$}
    \EndProcedure
  \end{algorithmic}
\end{algorithm}
Where $V_{wt}$ is the incrementor for wait time score. We take $V_{wt}$ as 1.
\subsection{Lexicographic VS IKEP Method of Tie Breaking}
In lexicographic tie breaking, the donors are either sequenced based on the name of patients or based on first come first serve. And the one which is placed earlier among the pairs, who form a tie is selected as the most preferred pair. The other way of tie breaking has been described in \cref{prioritizingPairs}.

Let \cref{tab:exampleCharacteristicTable_lexicovsikep} be another characteristic table with initial priorities as defined in the second last column. The last column describes the most preferred donors using lexicographic tie breaking versus IKEP stated tie breaking methodology.

\begin{sidewaystable}
\caption{Characteristic table for visualizing Lexicographic Vs IKEP method for tie breaking}
\label{tab:exampleCharacteristicTable_lexicovsikep}
\vspace{2 mm}
\centering
\begin{tabular}{c c c c c c c c c} \hline \\
Name  & Blood & HLA &Age &Kidney &PinCode &Societal Pref.(patient) &Init. &Top Pref \\ 
&Group&&& Size&&/Societal Dist.(donor) &Priority&(Lexi vs IKEP)\\ \hline
$P_1$ & $A^-$  & A1,B8,DR10,A3,B14,DR17 &45 &11 &496001 &$\emptyset$ &\multirow{2}{*}{$25$} &\multirow{2}{*}{$D_2$ vs $D_2$}\\
$D_1$ & $B^+$  & A2,B7,DR11,A10,B16,DR8 &30 &12 &496001 &$sd_1$\\
$P_2$ & $AB^+$ & A1,B8,DR17,A10,B16,DR8 &25 &11.5 &496001 &$\emptyset$ &\multirow{2}{*}{$29$} &\multirow{2}{*}{$D_1$ vs $D_5$}\\
$D_2$ & $O^-$  & A1,B8,DR17,A10,B16,DR8 &55 &11 &496001 &$sd_2$\\
$P_3$ & $A^-$  & A1,B8,DR17,A10,B16,DR8 &67 &12 &496001 &$\emptyset$ & \multirow{2}{*}{$5$} &\multirow{2}{*}{$D_2$ vs $D_2$} \\
$D_3$ & $A^+$  & A1,B8,DR17,A10,B16,DR8 &65 &11 &496001 & $sd_3$\\
$P_4$ & $AB^+$ & A1,B8,DR17,A10,B16,DR8 &30 &10 &496001 &$\emptyset$ & \multirow{2}{*}{$6$} &\multirow{2}{*}{$D_1$ vs $D_2$} \\
$D_4$ & $B^-$  & A1,B8,DR17,A10,B16,DR8 &66 &10 &490020 & $sd_2$\\
$P_5$ & $A^+$  & A1,B8,DR17,A10,B16,DR8 &60 &11.5 &496001 &$\emptyset$ & \multirow{2}{*}{$28$} &\multirow{2}{*}{$D_2$ vs $D_2$}\\
$D_5$ & $O^+$  & A1,B8,DR17,A10,B16,DR8 &55 &11.5 &496001 &$sd_1$\\ \hline
\end{tabular}
\end{sidewaystable}

\begin{figure}[H]
\centering
\includegraphics{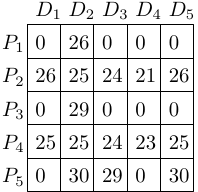}
\caption{Compatibility Matrix}
\label{IKEA_compatibilityMatrix_lexivsikep}
\end{figure}

\cref{IKEA_compatibilityMatrix_lexivsikep} is the compatibility matrix, calculated for \cref{tab:exampleCharacteristicTable_lexicovsikep}.

\begin{figure}
 \begin{subfigure}[b]{0.5\textwidth}
    \centering
     \includegraphics[width=8cm]{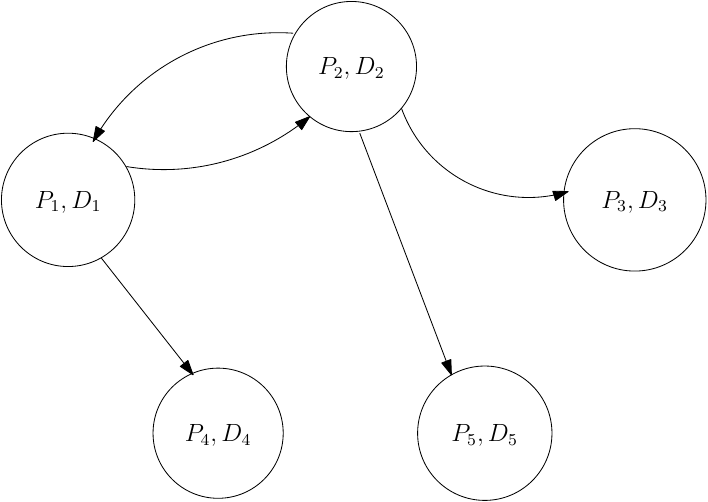}
     \caption{Lexicographical tie breaking}
     \label{fig:lexivsikepCompatibilityGraph11}
 \end{subfigure}
 \hspace{2cm}
 \begin{subfigure}[b]{0.5\textwidth}
     \centering
     \includegraphics[width=8cm]{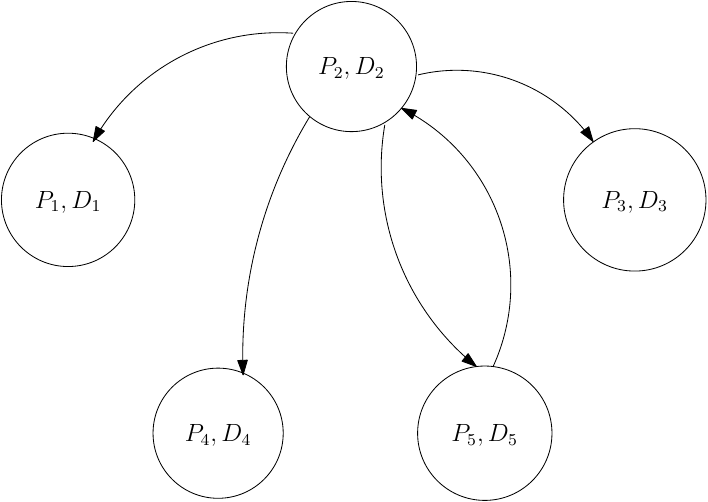}
     \caption{IKEP proposed tie breaking}
     \label{fig:lexivsikepCompatibilityGraph12}
 \end{subfigure}
 \caption{Compatibility Graph}
    \label{fig:lexivsikepCompatibilityGraph1}
\end{figure}

\cref{fig:lexivsikepCompatibilityGraph1} is the graphical representation of \cref{IKEA_compatibilityMatrix_lexivsikep}. \cref{fig:lexivsikepCompatibilityGraph11} is generated when the tie breaking method used is lexicographic where as, \cref{fig:lexivsikepCompatibilityGraph12} is generated when tie breaking is done using priority as described in \cref{prioritizingPairs}.

\begin{figure}
 \begin{subfigure}[b]{0.5\textwidth}
  \centering
     \includegraphics[width=8cm]{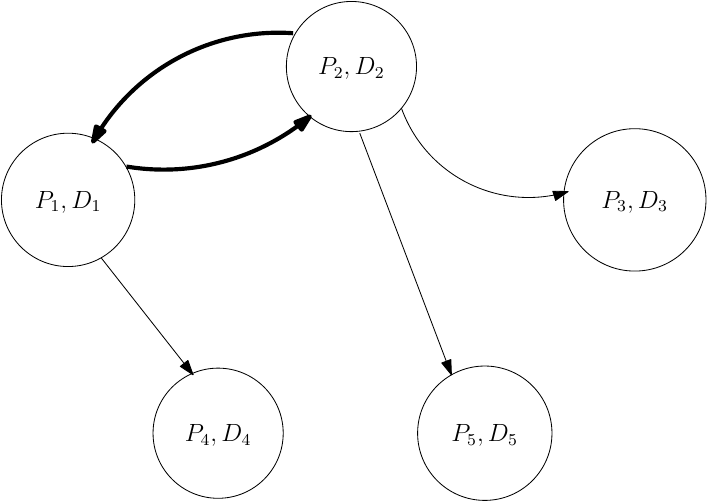}
     \caption{Lexicographical tie breaking}
     \label{fig:lexivsikepCompatibilityGraph21}
 \end{subfigure}
 \hspace{2cm}
 \begin{subfigure}[b]{0.5\textwidth}
     \centering
     \includegraphics[width=8cm]{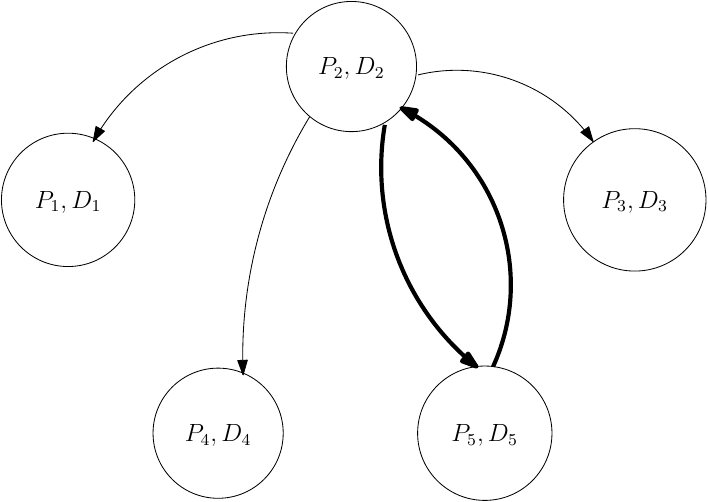}
     \caption{IKEP proposed tie breaking}
     \label{fig:lexivsikepCompatibilityGraph22}
 \end{subfigure}
 \caption{Compatibility Graph First Iteration}
    \label{fig:lexivsikepCompatibilityGraph2}
\end{figure}

The change in top preferences leads to the change in the graph and cycles present. \cref{fig:lexivsikepCompatibilityGraph2} shows the proposed exchange cycles in first iteration. 
\cref{fig:lexivsikepCompatibilityGraph21} proposes the exchange cycle as  $(P_1,D_1)$ $\longrightarrow$ $(P_2,D_2)$ $\longrightarrow$ $(P_1,D_1)$. Whereas, \cref{fig:lexivsikepCompatibilityGraph22} proposes the exchange cycle as $(P_2,D_2)$ $\longrightarrow$ $(P_5,D_5)$ $\longrightarrow$ $(P_2,D_2)$.

The cycles proposed in \cref{fig:lexivsikepCompatibilityGraph1} are then saved and the pairs participating in the cycles are removed out of the graph. 

\begin{figure}
 \begin{subfigure}[b]{0.5\textwidth}
     \centering
     \includegraphics[width=6cm]{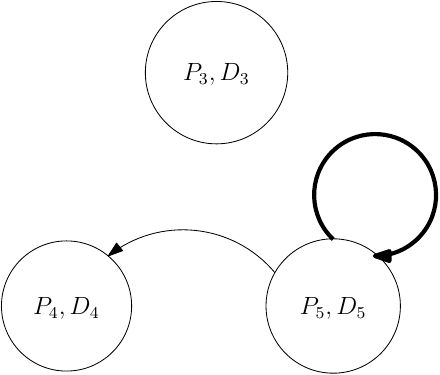}
     \caption{Lexicographical tie breaking}
     \label{fig:lexivsikepCompatibilityGraph31}
 \end{subfigure}
 \hspace{0.5cm}
 \begin{subfigure}[b]{0.5\textwidth}
     \centering
     \includegraphics[width=6cm]{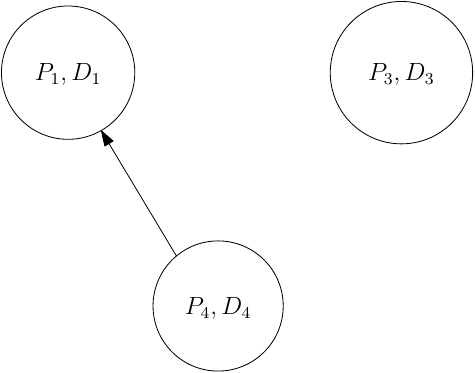}
     \caption{IKEP proposed tie breaking}
     \label{fig:lexivsikepCompatibilityGraph32}
 \end{subfigure}
 \caption{Compatibility Graph Second Iteration}
    \label{fig:lexivsikepCompatibilityGraph3}
\end{figure}

\cref{fig:lexivsikepCompatibilityGraph3} shows the new compatibility graph with the remaining pairs. \cref{fig:lexivsikepCompatibilityGraph31} and \cref{fig:lexivsikepCompatibilityGraph32} show the compatibility graphs of lexicographical and priority based tie breaking methods respectively. In the lexicographical tie breaking method, one more cycle possible is $(P_5,D_5)$ $\longrightarrow$ $(P_5,D_5)$. Whereas, no more cycles are possible in \cref{fig:lexivsikepCompatibilityGraph32}.

The final result of IKEP algorithm with lexicographical tie breaking mechanism is:
\[(P_1,D_1) \longrightarrow (P_2,D_2) \longrightarrow (P_1,D_1)\], \[(P_5,D_5) \longrightarrow (P_5,D_5)\]

and of IKEP algorithm with priority based tie breaking mechanism is:
\[(P_2,D_2) \longrightarrow (P_5,D_5) \]

Now we propose a way to deploy the results generated by IKEP.

\section{Deployment of exchanges}
\label{IKEP_deployment}
Due to the immense population of country, huge estimated participants in India, and prohibition of chains, encourages non-simultaneous cycles, which requires socialization to encourage trust. A priori, socialization becomes difficult when the number of people, financial structure, diversity and logistics challenged by geography comes into existence.
\subsection{Cycle Implementation}
Instead of following the common paradigm of conducting the procurement and transplantation simultaneously in the same hospital, which takes 2x surgical teams and operation theatre . IKEP takes out the kidneys simultaneously in the same hospital and would require 1x surgical teams and operation theatre. Hence, a cycle length could be increased to 2x. The execution of the cycles in the matching can be done by following steps.

\textbf{Step 1. }Donors travel to the location of recipients.

\textbf{Step 2. }Kidneys are procured from the donor at each center simultaneously. And operation theatre in all the centers act in a synchronised manner.

\textbf{Step 3. }Transplant the procured kidneys into the patients.

In the upcoming section, we simulate the IKEP run under different situations are discuss the results received.

\section{Simulation using Indian Data}
\label{simulationUsingIndianData}

In this section, all the "let us assume"(s) and algorithms proposed are taken into consideration and applied to the Indian specific data procured from \textbf{Indian Transplant Registry(ITR)}\footnote{Indian Transplant Registry(ITR) is a registry developed by Indian Society Of Organ Transplant on 24th April 1990 with a vision to generate and exchange transplant related information for promoting research.}.

In the simulations, we linearly regenerate data based on the demographics from ITR for 3 different simulation iterations, i.e., for pools of 100, 1000, 10000 pairs respectively. We don't take altruistic donors into considerations, because of prohibitory legislation.

\subsection{Outline Of Simulation}
\label{simulationOutline}
The simulation is performed as follows:
\begin{enumerate}

    \item Initially We visualise the effects of Indian influence on weights through the cardinality of pairs matched after running the IKEPA with and without taking SAS into consideration.
    
    \item We then see, how does the minority societal distributions get affected in terms of waiting time and probability of getting matched.
    
    \item After that, we simulate the IKEPA based on an idea discussed in Ashlagi et al. 2013\cite{kidneyExchangeInDynamicSparseHeterogenousPools}. Instead of running the IKEPA directly over a pool of 10000 pairs for say, we take smaller chunks of say, $10, 20, \dots100$ pairs, run IKEPA on it, and then add $10, 20, \dots100$ pairs in different simulation iterations respectively in order to see the effect of chunk matching and finding out an optimum value of chunk size and maximising the cardinality of patients matched.
    
    \item At the end, we simulate the enhanced form of IKEPA, i.e, E-IKEPA, as discussed in \cref{enhancedIKEPA}, and draw a picture of quantitative and qualitative improvement brought in by E-IKEPA.
    
\end{enumerate}

\subsection{Demographics}
The total number of live donors and patients around which, the data is oriented, are 383 and 195 respectively. 

\cref{tab:bloodGroupDistributions} represents the frequency distributions of donors and patients based on blood groups along with the rhesus factors. \cref{tab:ageGroupDistributions} shows us the distribution of patient-donor data based on 8 age groups. And at last but not the least, \cref{tab:stateWiseDistributions} gives us the patient-donor state wise distribution in India.

The distribution has been generated in such a way to clearly visualise the effects of differential distribution of participating pairs\footnote{In order to visualize the outcomes through the lenses of majority and minorities.}.

The societal distributions have been named as $sd_{index}$ in order to generalise the concept and can be implemented in many other non biological preference mechanisms.

\begin{table}
\caption{Blood group distributions}
\label{tab:bloodGroupDistributions}
\vspace{2 mm}
\small{\centerline{\begin{tabular}{l l c c} \hline
Blood & Rhesus & Donor Freq. & Recipient Freq. \\
Groups&Factor& as per data(\%)&as per data(\%)\\ \hline
\multirow{2}{*}{$A$} &$+ve$  &19.84 &28.21\\
    &$-ve$ &4.44 &5.64\\
\multirow{2}{*}{$B$} &$+ve$  &20.37 &24.10\\
    &$-ve$ &1.04 &1.54\\
\multirow{2}{*}{$AB$} &$+ve$  &1.04 &4.10\\
     &$-ve$ &0 &0\\
\multirow{2}{*}{$O$} &$+ve$  &50.91 &34.36\\
     &$-ve$ &2.35 &2.05\\
\hline
Total& &100 &100\\
\hline
\end{tabular}}}
\end{table}

\begin{table}
\caption{State Wise Patient-Donor distributions}
\label{tab:stateWiseDistributions}
\vspace{2 mm}
\small{\centerline{\begin{tabular}{l c c} \hline
\multirow{2}{*}{State} & Donor Freq. & Recipient Freq. \\
& as per data(\%)&as per data(\%)\\ \hline
Andaman Nicobar&3.92 &4.10\\
Andhra Pradesh&1.04 &0.51\\
Bihar&0.78&1.03\\
Delhi &3.13 &7.18\\
Haryana&1.57&0.51\\
Himachal Pradesh&0.26&3.59\\
International &17.49&0.51\\
Jammu$\And$Kashmir&0.26&4.10\\
Jharkhand&0.26&0.51\\
Karnataka &40.47 &19.49\\
Kerala&0.26&0\\
Madhya Pradesh&1.57&2.56\\
Maharashtra&1.31&0.51\\
Manipur&0.26&0.51\\
Orissa&0.26&0.51\\
Rajasthan&1.31&1.54\\
Tamil Nadu &23.76 &48.21\\
Uttar Pradesh&1.82&4.10\\
West Bengal&0.26&0.51\\
\hline
Total &100 &100\\
\hline
\end{tabular}}}
\end{table}

\begin{table}
\caption{Age group distributions}
\label{tab:ageGroupDistributions}
\vspace{2 mm}
\small{\centerline{\begin{tabular}{l c c} \hline
Age & Donor Freq. & Recipient Freq. \\
Groups& as per data(\%)&as per data(\%)\\ \hline
0-10  &0 &2.05\\
11-20 &1.04 &5.64\\
21-30 &22.19 &24.10\\
31-40 &32.11 &21.03\\
41-50 &23.76 &18.97\\
51-60 &15.67 &21.03\\
61-70 &4.70 &7.18\\
$\>$70 &0.52 &0\\
\hline
Total &100 &100\\
\hline
\end{tabular}}}
\end{table}

\subsection{Construction Of Data}
For reducing the complexity of the simulations, we take the following assumptions into considerations.
\begin{assumption}
\label{assumption1}
All the statistical data retrieved, are linear with the increase in $\mathbf{card}(pairs)$.
\end{assumption}

\begin{assumption}
\label{assumption2}
Societal distributions are equally distributed all over the nation.
\end{assumption}

Since there are very less kidney registry data available in India, we had to multiply the data in order to simulate a scenario of 100 to 10,000 pairs.

We have procured data from two sources in order to simulate the IKEPA described in \cref{IKEP_algorithms}.

After procuring data from the following registries, we calculate the percen2tage distribution of participants categorically. And based on the percentage distributions, regenerated data according to the cardinality specified. The generated data is then shuffled. Based on this shuffled data, the final database of patient donor data is created.

For example, in a data consisting of 4 patients of blood group A, 2 patients of blood group AB, 3 patients of blood group O and 1 patients of blood group B. And let the cardinality specified be 1000. So the generating algorithm will take a vector of length 1000, fill the first $\frac{4}{10} \times 1000$ items with blood group A and similarly the others. The decimals are rounded off. 

After the vector being generated, all the values are shuffled stochastically and then used in the database \footnote{The database used for simulation can be found in github repository for \href{https://github.com/arghyabandyopadhyay/Indian-Kidney-Exchange-Program}{Indian Kidney Exchange Program}} generation.

For societal distribution and the societal preferences are generated entirely randomly\footnote{Since $sdPref$ depends on historical events of disharmony between $sd$(s), it has been taken randomly.} for the first, third and the fourth part of simulation as described in the outline of simulations in \cref{simulationOutline}. 

\subsection{Discussion}
Here, we discuss all the results generated by the simulations performed as stated in \cref{simulationOutline}.
\subsubsection{Notations}
Let, $pair^t$ is the number of pairs present in the pool during the time of IKEPA run., \[pair^m= \mathbf{card}(pairsMatched)\] is the number of pairs which got matched during the IKEPA run., \[L_{max}^{'}= [3, 10, 100, 1000, 10000]\] is the set of all the $L_{max}$ taken into consideration for simulation according to the cardinality of participating pairs. And, \[F^{*'} = [0, 5, 10, 15, 20, 25, 30]\] is the set of all $F^*$ taken into consideration for simulation.
\subsubsection{Influence Of Societal Acceptance Score}
\label{influenceOfSocietalAcceptanceScoreSimulation}
In the upcoming figures, we show the $pair^m$ trends for every edge filter $F^* \in F^{*'}$ in a pool of 100, 1000, 10000 patient-donor pairs respectively.

\begin{figure}
\centering
\includegraphics[width=8cm]{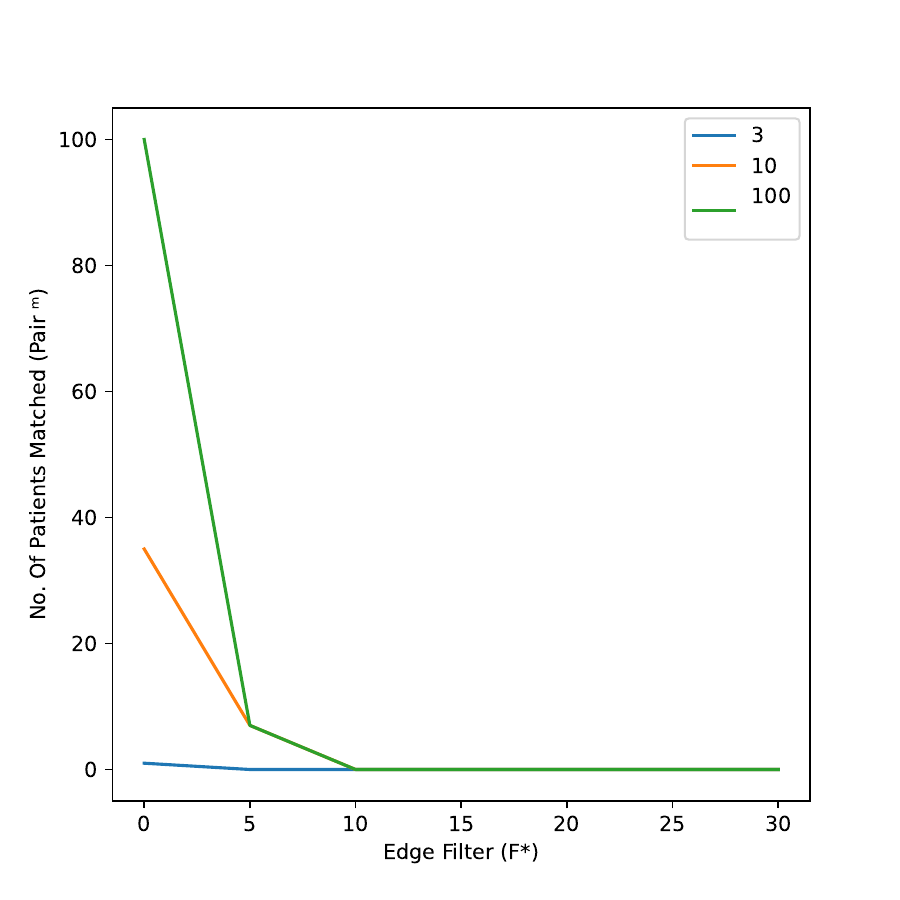}
\caption{$pair^m$ vs $F^*$ $\forall L_{max} \in L_{max}^{'}$ in a pool of 100 pairs influenced  by SAS}
\label{graph:patientsMatched_With_sample100}
\end{figure}

\cref{graph:patientsMatched_With_sample100} plots the line chart of $pair^m$ using different $F^*$ as described earlier. We observe that, $pair^m$ is very less and in the upcoming figures, it increases with the increase in number of pairs.

\begin{figure}
\centering
\includegraphics[width=8cm]{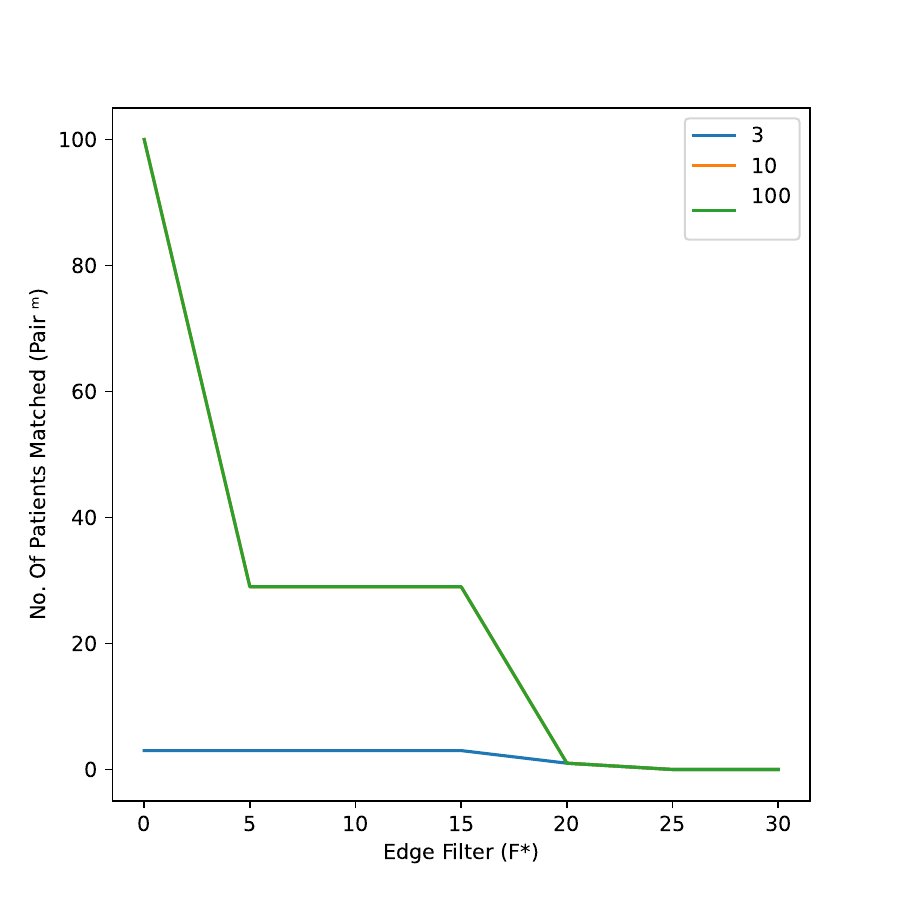}
\caption{$pair^m$ vs $F^*$ $\forall L_{max} \in L_{max}^{'}$ in a pool of 100 pairs without the influence of SAS}
\label{graph:patientsMatched_WithoutSAS_sample100}
\end{figure}

Whereas, in \cref{graph:patientsMatched_WithoutSAS_sample100} plots the line chart of $pair^m$ using different $F^*$ as described earlier but without the influence of SAS. The difference can be inferred from the contradictions with the previous observation. And the exchanges which gets discarded with the influence of SAS, represent the societally unacceptable kidneys by respective participating pairs.

The $pair^m$ in the \cref{graph:patientsMatched_WithoutSAS_sample100} for $F^*= 15$, 20 unlike in the range 0-2 as seen in the \cref{graph:patientsMatched_With_sample100}.

\begin{figure}
\centering
\includegraphics[width=8cm]{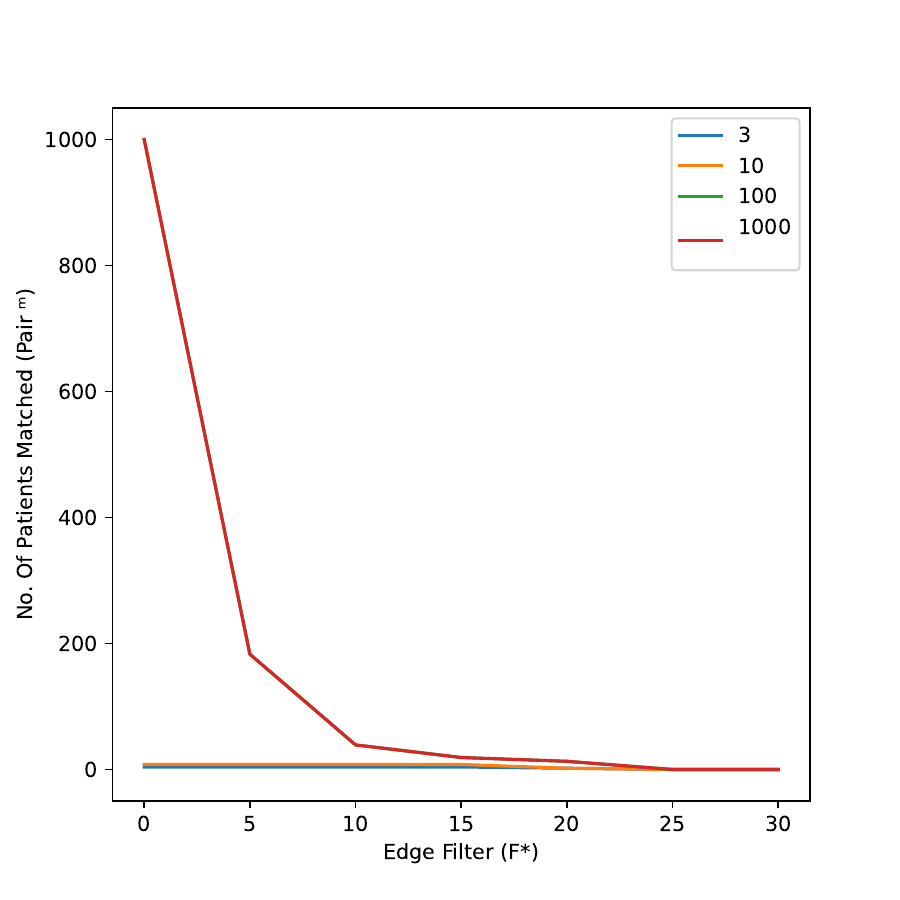}
\caption{$pair^m$ vs $F^*$ $\forall L_{max} \in L_{max}^{'}$ in a pool of 1000 patients influenced by SAS}
\label{graph:patientsMatched_With_sample1000}
\end{figure}

\cref{graph:patientsMatched_With_sample1000}, when compared with \cref{graph:patientsMatched_With_sample100}, it can be noticed that, $pair^m$ is more. It can also be observed from all the graphs referred yet, that, $pair^m$ reduces to 0 at $F^*= 25$.

\begin{figure}
\centering
\includegraphics[width=8cm]{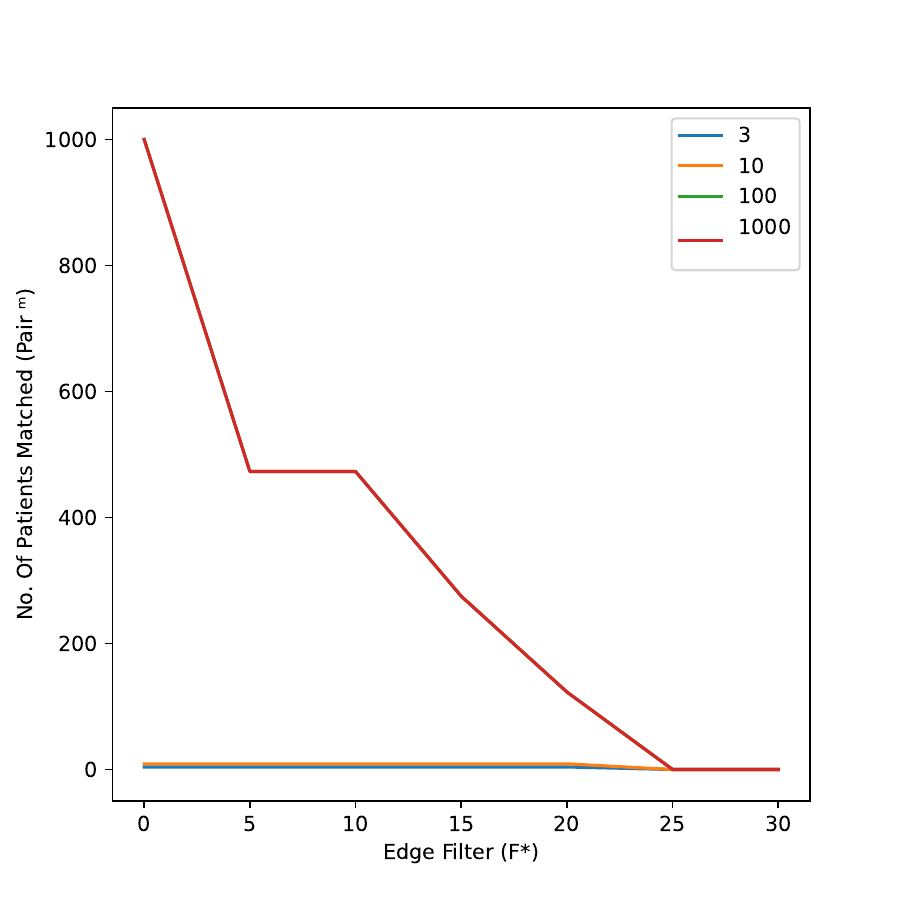}
\caption{$pair^m$ vs $F^*$ $\forall L_{max} \in L_{max}^{'}$ in a pool of 1000 pairs without the influence of SAS}
\label{graph:patientsMatched_WithoutSAS_sample1000}
\end{figure}

And when we try to simulate the same data set without the influence of societal acceptance score as shown in \cref{graph:patientsMatched_WithoutSAS_sample1000}, it is again observed that there are a significant amount of potentially rejectable pairs.

\begin{figure}
\centering
\includegraphics[width=8cm]{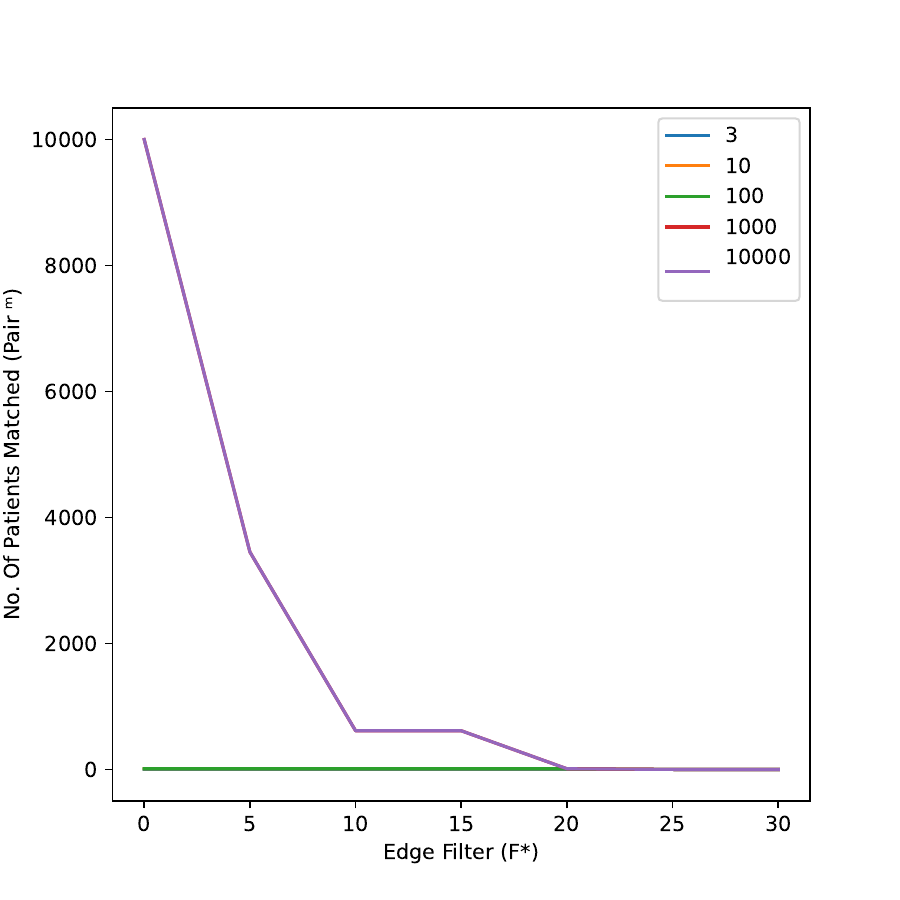}
\caption{$pair^m$ vs $F^*$ $\forall L_{max} \in L_{max}^{'}$ in a pool of 10000 patients influenced by SAS}
\label{graph:patientsMatched_With_sample10000}
\end{figure}

Now we simulate for a data set of 10,000 pairs. \cref{graph:patientsMatched_With_sample10000} shows the trend followed by $pair^m$, it reduces to 0 at $F^*= 20$. Since the data is generated according to the \cref{assumption1}, the graph generated are more or less depict similar trends.

\begin{figure}
\centering
\includegraphics[width=8cm]{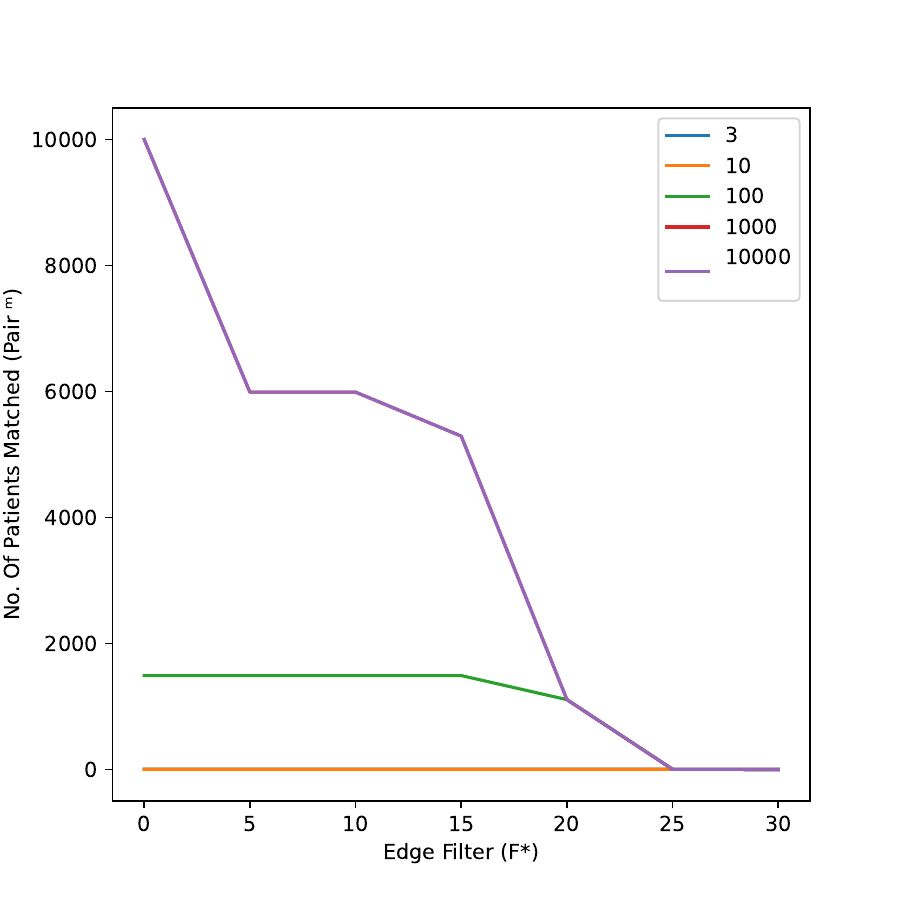}
\caption{$pair^m$ vs $F^*$ $\forall L_{max} \in L_{max}^{'}$ in a pool of 10000 pairs without the influence of SAS}
\label{graph:patientsMatched_WithoutSAS_sample10000}
\end{figure}

It can also be observed that, with the increase in $F^*$, $pair^m$ decreases on an average. 

For $F^*= 0,\ pair^m = \mathbf{card}($pairs$)$ 

For $F^*= 30,\ pair^m = 0$

We can also infer from all of the graphs presented above, that, the $pair^m$ is maximum when the $L_{max}= \mathbf{card}(pairs)/10$. The percentage of $pair^m$ for a rational value of $F^*= 5 \And L_{max}= \mathbf{card}(pairs)/10\ \forall\ \mathbf{card}(pairs)\in\{100,1000,10000\}$ is approximately 8$\%,$ 18$\%$, 35$\%$ respectively. Hence, with the increase in number of patients in the pool, improves the percentage of pairs getting matched.

\subsubsection{Quantum Matching In IKEP}
\label{quantumMatchingInIKEPSimulation}
We now, re-simulate a patient-donor data of 10000 pairs used in \cref{graph:patientsMatched_With_sample10000}, using the method proposed in Ashlagi et al. 2013\cite{kidneyExchangeInDynamicSparseHeterogenousPools}. We simulate using different quantum sizes of patient-donor pairs, i.e, 1000, 100, 10 and 1, i.e, pairs arrive into the pool at these particular rates in different IKEPA run. After that, we compare the graphs generated with the one which is generated in \cref{graph:patientsMatched_With_sample10000}.

\begin{figure}
\centering
\includegraphics[width=8cm]{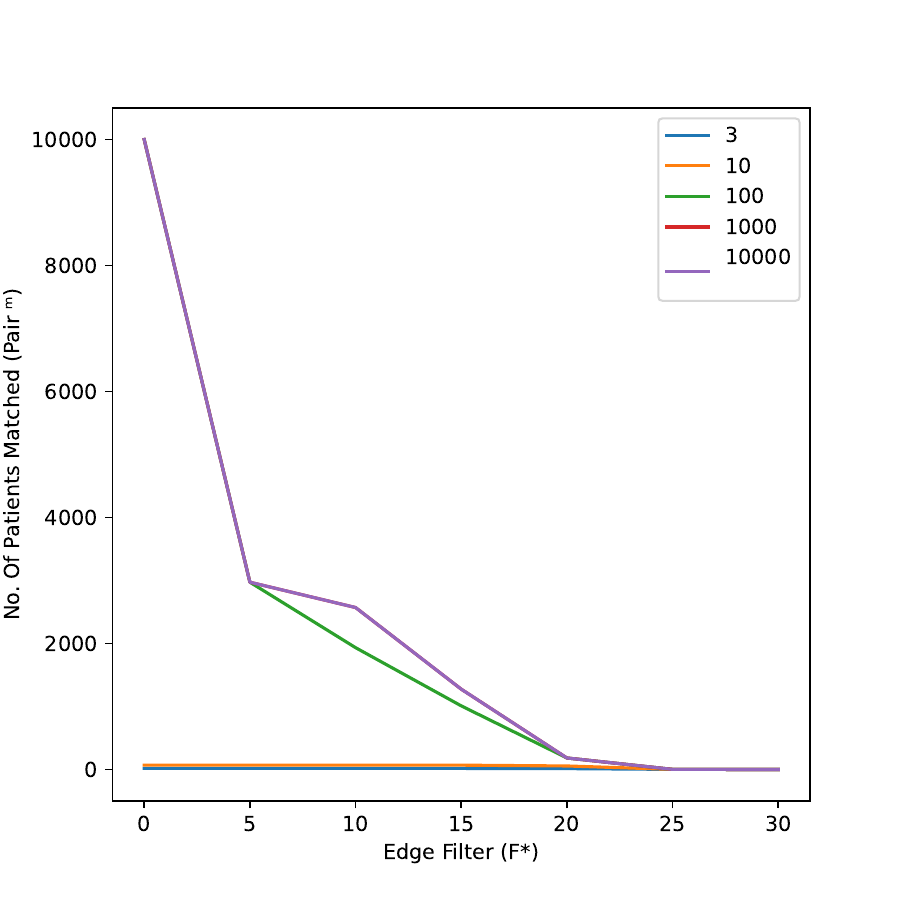}
\caption{$pair^m$ vs $F^*$ $\forall L_{max} \in L_{max}^{'}$ in a pool of 10000 pairs with arrival rate of 1000 at a time.}
\label{graph:PatientsMatched_Dynamic_With_sample10000_Result_maxCard_1000}
\end{figure}

\cref{graph:PatientsMatched_Dynamic_With_sample10000_Result_maxCard_1000} describes the trend followed by the $pair^m$, when we initially run the IKEPA on a pool of first 1000 pairs selected out of a pool of 10000 pairs. All the pairs that get matched up in the first iteration are removed from the pool, others remain in the pool and then, next 1000 pairs are introduced in to the existing pool. Again, IKEPA is run on this pool of pairs. This is repeated again and again, until and unless, all the 10000 pairs have already been added up to the pool. For $F^*= 10$, the $pair^m$ increases from 5$\%$ in \cref{graph:patientsMatched_With_sample10000} to 27$\%$ which is a massive improvement.


\begin{figure}
\centering
\includegraphics[width=8cm]{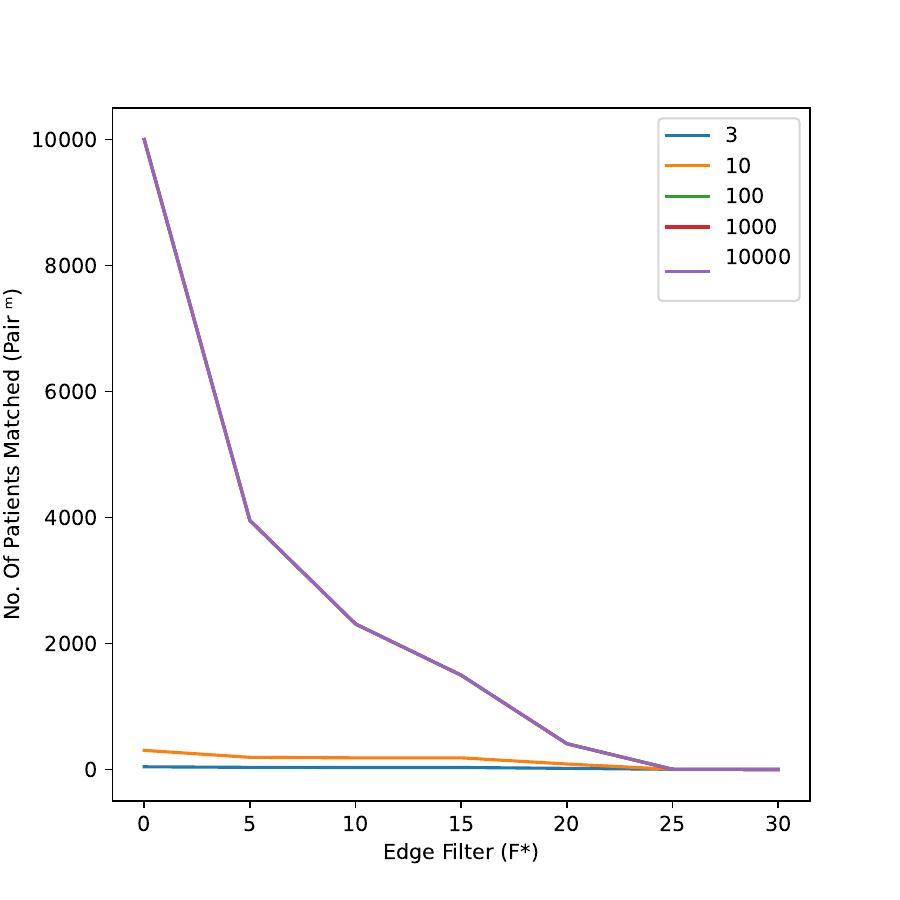}
\caption{$pair^m$ vs $F^*$ $\forall L_{max} \in L_{max}^{'}$ in a pool of 10000 pairs with arrival rate of 100 at a time.}
\label{graph:PatientsMatched_Dynamic_With_sample10000_Result_maxCard_100}
\end{figure}

Similarly, we run the IKEPA on a pool which starts with 100 patient-donor pairs and 100 pairs keep on adding up after each algorithm run. We can observe that, for $F^*= 5$, the $pair^m$, which is 38$\%$, is greater than what it is in \cref{graph:PatientsMatched_Dynamic_With_sample10000_Result_maxCard_1000} opposite for $F^*= 10$. And when compared with, \cref{graph:patientsMatched_With_sample10000}, for $F^*= 5$, the $pair^m$ is also greater in \cref{graph:PatientsMatched_Dynamic_With_sample10000_Result_maxCard_100}.

With the decrease in the quantum size, we can also observe that, for $F^*= 20$, $pair^m$ increases slightly.

Now, we run the IKEPA on a pool which starts with 10 patient-donor pairs and 10 pairs keep on adding up after each algorithm run. It is found that, for $F^*= 5$, the $pair^m$ increases upto 53$\%$ of the pairs present in the pool at any given time, whereas, it is observed that, $pair^m$ decreases for $F^*= 10$, when compared to previous simulations.

\begin{figure}
\centering
\includegraphics[width=8cm]{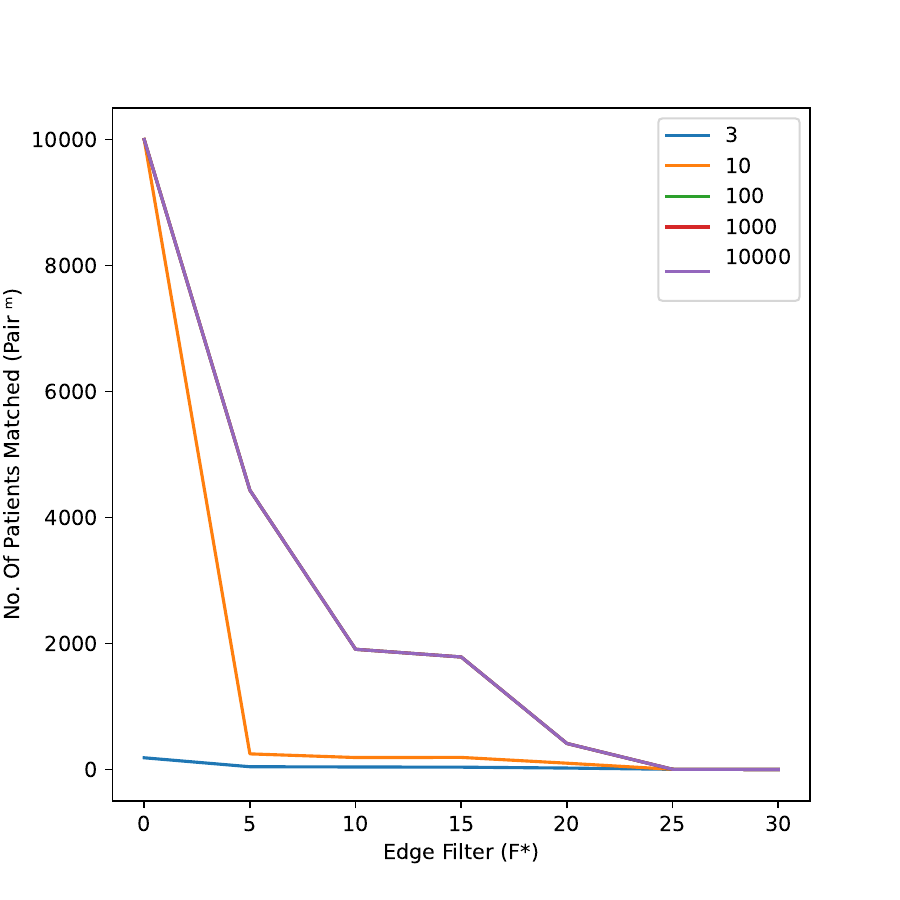}
\caption{$pair^m$ vs $F^*$ $\forall L_{max} \in L_{max}^{'}$ in a pool of 10000 pairs with arrival rate of 10 at a time.}
\label{graph:PatientsMatched_Dynamic_With_sample10000_Result_maxCard_10}
\end{figure}

And at last, we run the IKEPA on a pool which starts with only 1 patient-donor pair and 1 pair keep on adding up after each algorithm run. This resembles to the greedy approach described in \cref{dynamicMatchingBackground}. In this mechanism, at $F^*=5$, approximately 42\% of the pairs are matched, and $F^*=10$ matches 30\% of the pairs in the pool. Hence, $pair^m$ reduces when, arrival rate is set to 1, which is one of the short comings of greedy algorithm.

\begin{figure}
\centering
\includegraphics[width=8cm]{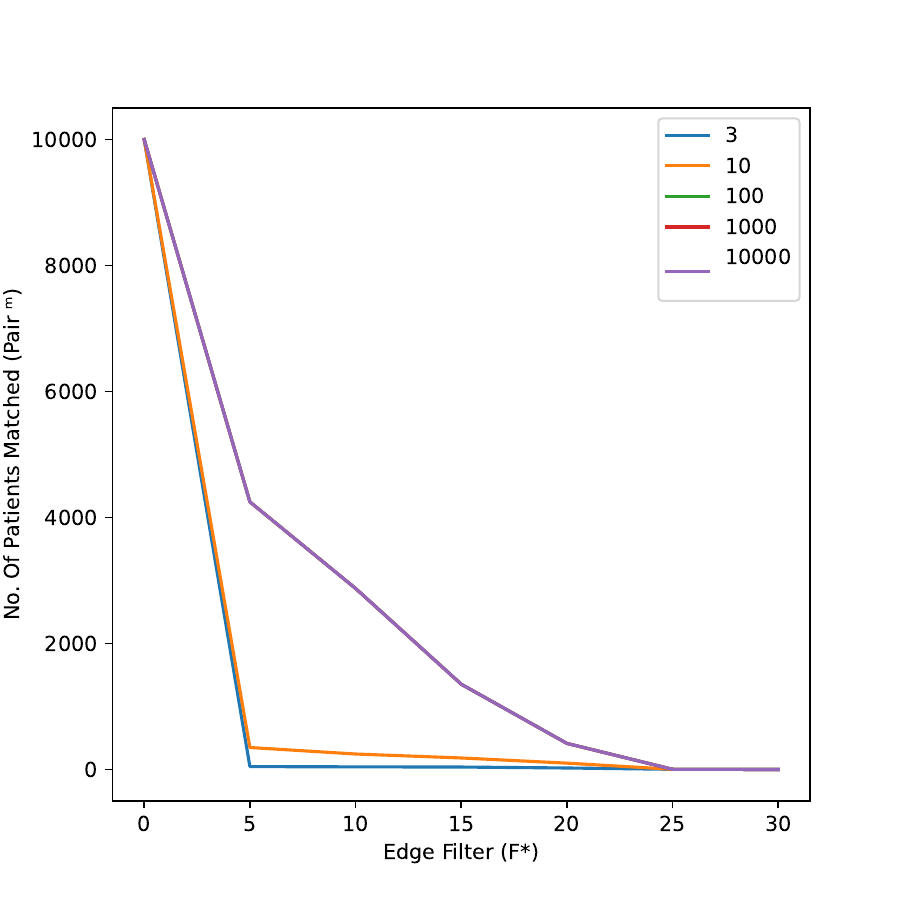}
\caption{$pair^m$ vs $F^*$ $\forall L_{max} \in L_{max}^{'}$ in a pool of 10000 pairs with arrival rate of 1 at a time.}
\label{graph:PatientsMatched_Dynamic_With_sample10000_Result_maxCard_1}
\end{figure}

The results generated by the experimental Quantum matching, points out a fact, that the number of pairs matched by IKEP increases with the decrease in chunk sizes. The optimal chunk size is found to be 10.

The reason behind this is, the chances of forming a deadlock increases with larger pools. This problem is then solved by the enhanced IKEPA, termed as E-IKEPA as described in \cref{enhancedIKEPA}.

\subsubsection{Significant increase in matched pairs by E-IKEPA}
\label{eikepaSimulationResults}
Now, we re-run the simulation done in \cref{influenceOfSocietalAcceptanceScoreSimulation} but with E-IKEPA in the place of IKEPA and compare the result with those generated in \cref{influenceOfSocietalAcceptanceScoreSimulation}, \cref{quantumMatchingInIKEPSimulation}.

\begin{figure}
\centering
\includegraphics[width=8cm]{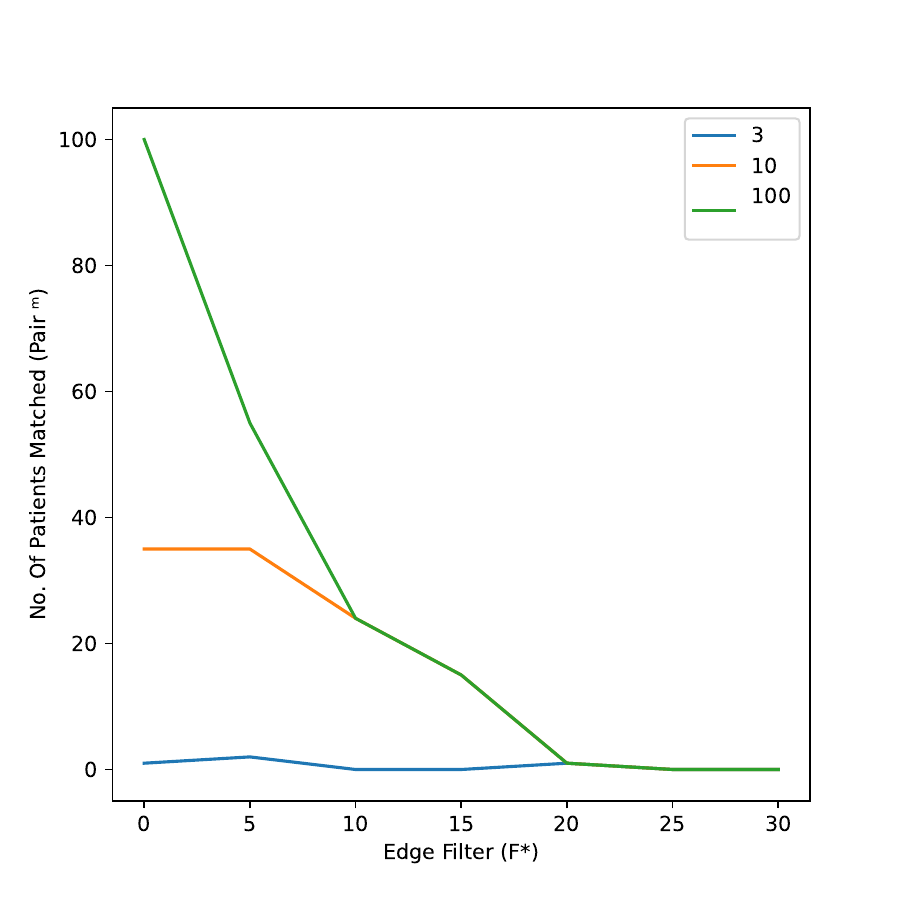}
\caption{$pair^m$ vs $F^*$ $\forall L_{max} \in L_{max}^{'}$ in a pool of 100 pairs using E-IKEPA.}
\label{graph:PatientsMatched_sample100_enhanced}
\end{figure}

\cref{graph:PatientsMatched_sample100_enhanced} plots the $pair^m$ in a pool of 100 pairs, using E-IKEPA. When compared with \cref{graph:patientsMatched_With_sample100}, it can be observed that, at $F^*=10$, the difference in $pair^m$ by E-IKEPA is approximately 25-27 pairs, using $L_{max}=10$. 

\begin{figure}
\centering
\includegraphics[width=8cm]{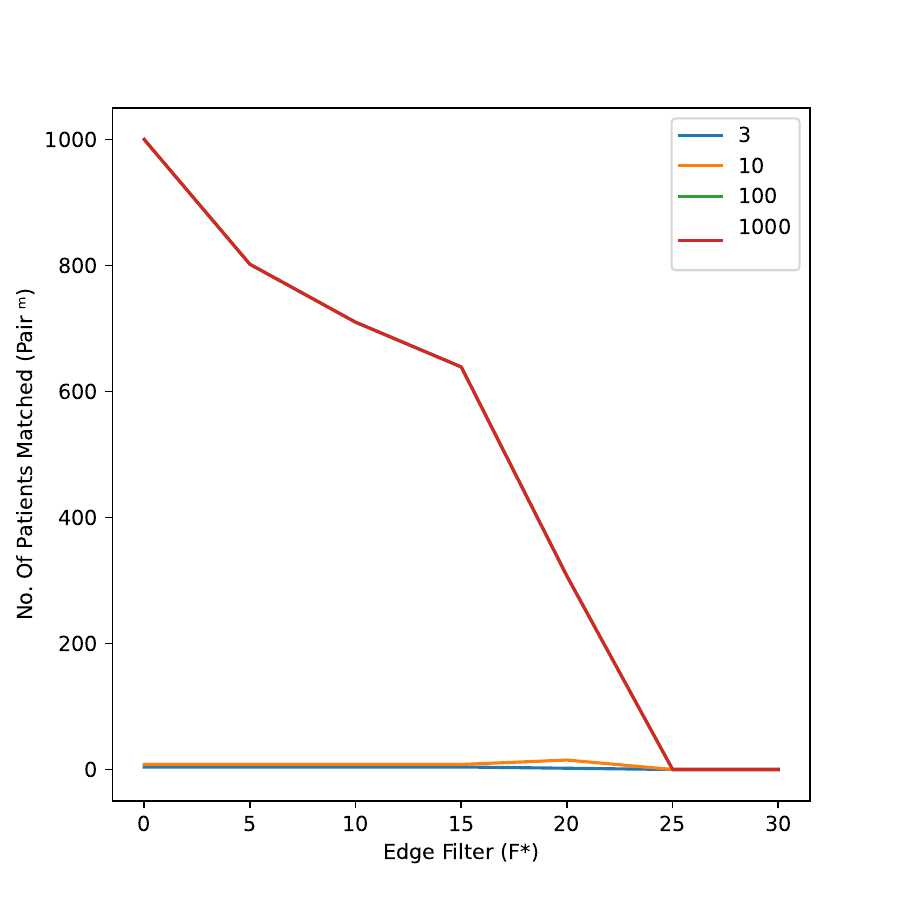}
\caption{$pair^m$ vs $F^*$ $\forall L_{max} \in L_{max}^{'}$ in a pool of 1000 pairs using E-IKEPA.}
\label{graph:PatientsMatched_sample1000_enhanced}
\end{figure}

Similarly, but at a larger scale, \cref{graph:PatientsMatched_sample1000_enhanced} matches near about 700 pairs at $F^*=10$, unlike \cref{graph:patientsMatched_With_sample1000}, which matches only 10-20 pairs. The differences are quite stunning. Even if $F^*$ is set to 20\footnote{According to the compatibility quantizations applied, 20 is a pretty good score to have.}, E-IKEPA is able to match 40-50\% of the patients present in the pool unlike 5-10\% matched by IKEPA at $L_{max}=10\%$ of $\mathbf{card}(pairs)$.

\begin{figure}
\centering
\includegraphics[width=8cm]{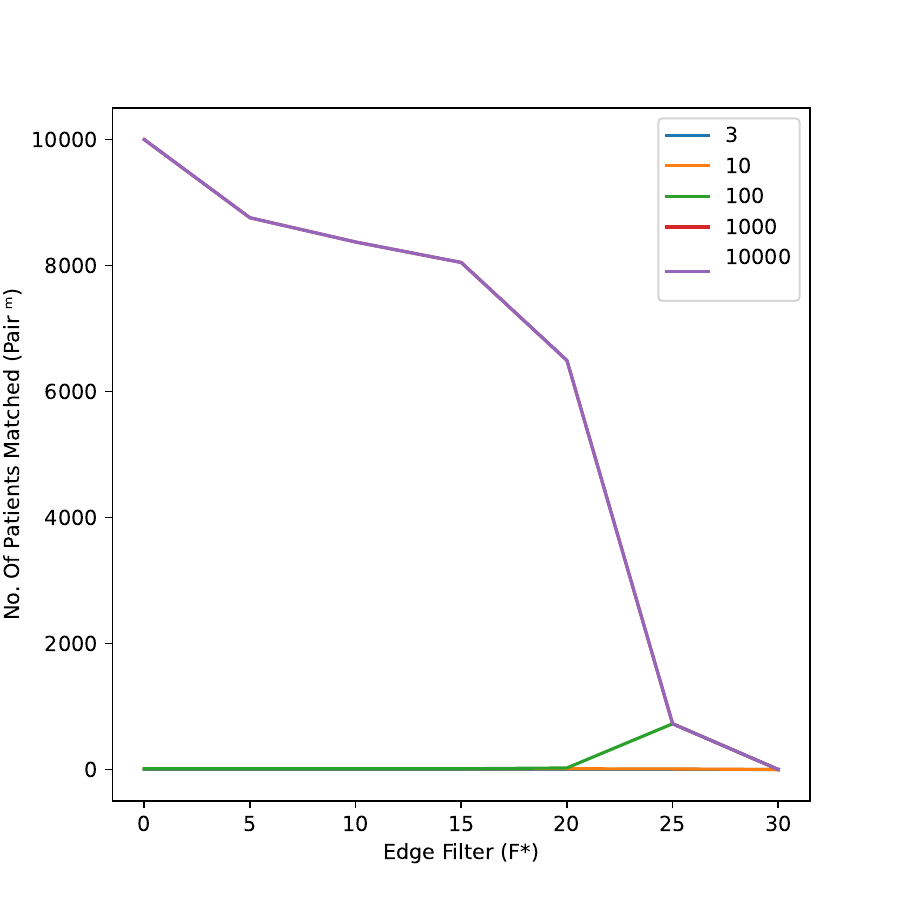}
\caption{$pair^m$ vs $F^*$ $\forall L_{max} \in L_{max}^{'}$ in a pool of 10000 pairs using E-IKEPA.}
\label{graph:PatientsMatched_sample10000_enhanced}
\end{figure}

Now, we compare \cref{graph:PatientsMatched_sample10000_enhanced} to \cref{graph:patientsMatched_With_sample10000}. Again taking two $F^*$ for reference, at $F^*=10$, the $pair^m$ by E-IKEPA is 8700 approximately, whereas IKEPA only matches, 500-600 pairs. And at $F^*=20$, the $pair^m$ appears to be 7000 by E-IKEPA vs 0 by IKEPA.

Even when we compare the results generated by E-IKEPA on a pool of 10000 pairs, to the result generated by running IKEPA, on the concept of quantum matching, discussed in \cref{quantumMatchingInIKEPSimulation}. The best result in quantum matching was generated by using arrival rate of 10, which in turn is found to be much more inferior than the result generated by E-IKEP.

The above observations show that how efficient, E-IKEPA is in finding huge amount of exchanges, relevant to Indian context.

\subsection{System Details}
All the simulations are run in an Ubuntu 20.04 operating system running on a Dell Laptop with 16GB Random Access Memory, 480GB Solid State Drive and i5 6th generation processor.

\section{Limitations of IKEP}
\label{limitationsOfIkep}
Most of the limitations of IKEP is either due to the literacy or due to legal restrictions\footnote{As described in \cref{legalFrameworkInIndia}.}. And hence can only be solved by government reforms and actions.
\subsection{Anonymity}
Since the pairs need to get themselves a permission to go for a transplant from the gazetted officers, and IKEP will also require to keep a track of the data, so that it could prove the legitimacy of the patient-donor pair in times of audits by government. Anonymity can only be conserved to a particular extent only.

\subsection{Prohibition of Altruistic Donors}
According to the laws in India for organ donations, it is illegal to donate organ without an intended recipient. This prohibition, indirectly removes the presence of chains in the KEPs. It has been observed that, KEPs without chains would eventually increase the average waiting time for the under demanded pairs and leads to accumulation. This is a serious limitation to IKEP, and proper amendments are needed to legalize it.

\subsubsection{Allowing chains will be revolutionary}
Let there be 5 pairs $(P_1,D_1) \dots (P_5,D_5)$. And the characteristics described in the \cref{tab:exampleCharacteristicTable1}. We have assumed for the ease of solution, that every patient don't have a preference on societal distributions.

\begin{sidewaystable}
\caption{Example participant details.}
\label{tab:exampleCharacteristicTable1}
\vspace{2 mm}
\centering
\begin{tabular}{c c c c c c c c c} \hline \\
Name  & Blood & HLA &Age &Kidney &PinCode &Societal Pref.(for patient) &Initial &Top \\ 
&Group&&&Size&&/Societal Dist.(for donor) &Priority&Pref.\\ \hline
$P_1$ & $A^-$  & A1,B8,DR10,A3,B14,DR17 &45 &11 &496001 &$\emptyset$ &\multirow{2}{*}{$5$} &\multirow{2}{*}{$D_2$} \\
$D_1$ & $B^+$  & A2,B7,DR11,A10,B16,DR8 &30 &12 &496001 &$sd_1$ \\
$P_2$ & $AB^+$ & A1,B8,DR10,A2,B7,DR11 &25 &10 &490020 &$\emptyset$ &\multirow{2}{*}{$2$} &\multirow{2}{*}{$D_3$} \\
$D_2$ & $O^-$  & A1,B8,DR10,A10,B16,DR8 &55 &11.5 &496001 &$sd_2$  \\
$P_3$ & $A^+$  & A1,B8,DR17,A10,B16,DR8 &67 &12 &496001 &$\emptyset$ &\multirow{2}{*}{$1$} &\multirow{2}{*}{$D_5$} \\
$D_3$ & $AB^+$  & A1,B8,DR17,A10,B16,DR8 &25 &11 &496001 &$sd_1$  \\
$P_4$ & $AB^+$ & A1,B8,DR17,A10,B16,DR8 &30 &10 &496001 &$\emptyset$ &\multirow{2}{*}{$6$} &\multirow{2}{*}{$D_4$} \\
$D_4$ & $B^-$  & A1,B8,DR17,A10,B16,DR8 &27 &12 &496001 &$sd_3$  \\
$P_5$ & $A^+$  & A1,B8,DR17,A10,B16,DR8 &60 &11.5 &496001 &$\emptyset$ &\multirow{2}{*}{$10$} &\multirow{2}{*}{$D_5$} \\
$D_5$ & $O^+$  & A1,B8,DR17,A10,B16,DR8 &55 &11.5 &496001 &$sd_2$  \\  \hline
\end{tabular}
\end{sidewaystable}

\begin{figure}[h]
\centering
\includegraphics{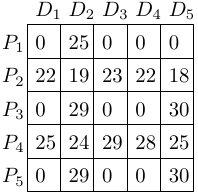}
\caption{Compatibility Matrix}
\label{IKEA_compatibilityMatrix1}
\end{figure}

According to the algorithm stated, the compatibility matrix W generated is shown in \cref{IKEA_compatibilityMatrix1}.

\begin{figure}[h]
\centering
\includegraphics[width=8cm]{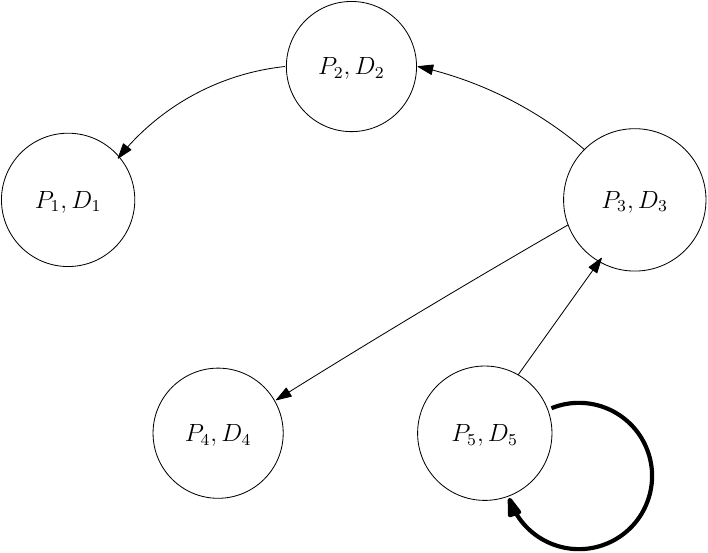}
\caption{Compatibility Graph}
\label{IKEA_compatibilityGraph1}
\end{figure}

The graph representation of the compatibility matrix is shown in \cref{IKEA_compatibilityGraph1}

Let us assume the priorities\footnote{It is actually assigned and modified with time according to various time dependent characteristics like results of previous IKEPA runs, arrival rate of pairs.} assigned to the patients be as described in the \cref{tab:exampleCharacteristicTable1}. And the top preferences are derived from the compatibility matrix.

\begin{figure}[h]
\centering
\includegraphics[width=8cm]{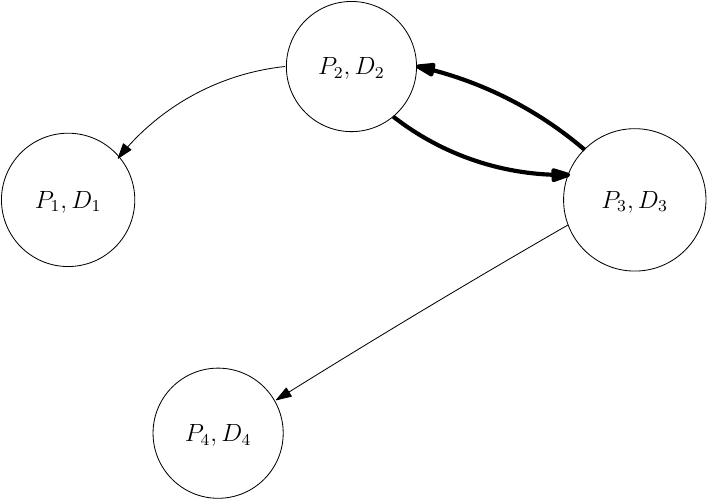}
\caption{Compatibility Graph in Second Iteration}
\label{IKEA_compatibilityGraph1_secondIteration}
\end{figure}

In the above example, not more than 1 pair cycles can be formed. Hence, all the remaining patients will keep on waiting until the next IKEPA run. And within the waiting time, there does exists a huge possibility of loosing the patient. Introducing chains in India does have much more pros than the cons. According to the priority list, the first exchange will provide kidney from donor $D_5$ to patient $P_5$. And after that no cycles can be formed in the first iteration. The compatibility graph for the second iteration is shown in \cref{IKEA_compatibilityGraph1_secondIteration}.

\begin{figure}[h]
\centering
\includegraphics[height=4.5cm]{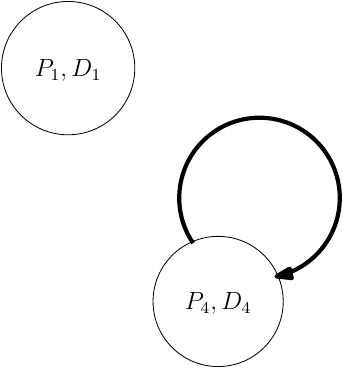}
\caption{Compatibility Graph in Third Iteration}
\label{IKEA_compatibilityGraph1_thirdIteration}
\end{figure}

After removing the $(P_2,D_2)$ and $(P_3,D_3)$ from the graph and exchange being carried out. The new compatibility graph would be like \cref{IKEA_compatibilityGraph1_thirdIteration}. And in \cref{IKEA_compatibilityGraph1_thirdIteration} it can be seen that, $P_4$ prefers $D_4$ and hence after entire completion of the IKEPA, $(P_1,D_1)$ remains unmatched.

Hence in order to increase the number of pairs getting positive outcomes, there are two ways.

\begin{itemize}
    
    \item \textbf{Increasing number of participants by waiting:} Let us assume a patient donor pair $(P_6,D_6)$ with \textbf{priority 4} which arrives at a particular point of time. The properties of the new pairs are given in \cref{tab:extendedCharacteristicTable}. And for ease of understanding, let the societal preference be $\emptyset$ and the societal distribution be $sd_1$.
    
    \begin{table}[htbp]
    \caption{Details of additional pairs}
    \label{tab:extendedCharacteristicTable}
    \vspace{2 mm}
    \small{\centerline{\begin{tabular}{c c c c c c} \hline
    Name  & ABO & HLA &Age &Size &PinCode \\ \hline
    $P_6$ & $B^+$  & A2,B7,DR11, &25 &11 &496001 \\
            &   & A10,B16,DR8    &   &   &   \\
    $D_6$ & $A^+$  & A1,B8,DR17, &67 &11.5 &496001 \\
            &   &A10,B16,DR8    &   &   &   \\\hline
    \end{tabular}}}
    \end{table}
    
    the new compatibility matrix W generated is shown in \cref{IKEA_newCompatibilityMatrix}.

    \begin{figure}[H]
    \centering
    \includegraphics{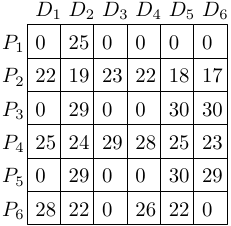}
    \caption{Extended Compatibility Matrix}
    \label{IKEA_newCompatibilityMatrix}
    \end{figure}
    
    The new graph representation of the compatibility matrix is shown in \cref{IKEA_compatibilityGraphWithMorePairs}. In this graph, the first iteration provides $P_5$ the kidney from $D_5$ which in turn modifies the graph into \cref{IKEA_compatibilityGraphWithMorePairs_secondIteration}.
    
    \begin{figure}[H]
    \centering
    \includegraphics[width=8cm]{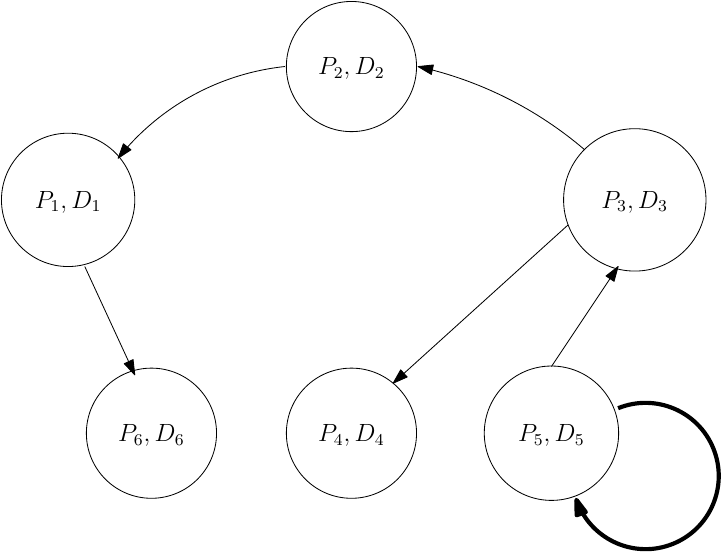}
    \caption{Compatibility Graph With More Pairs}
    \label{IKEA_compatibilityGraphWithMorePairs}
    \end{figure}
    
    In the second iteration, the cycle goes like $(P_1,D_1) \longrightarrow (P_2,D_2) \longrightarrow (P_3,D_3) \longrightarrow (P_4,D_4) \longrightarrow (P_1,D_1)$. After removal of all the participating vertices, the modified compatibility graph for third iteration is shown in \cref{IKEA_compatibilityGraphWithMorePairs_thirdIteration}.
    
    \begin{figure}[H]
    \centering
    \includegraphics[width=8cm]{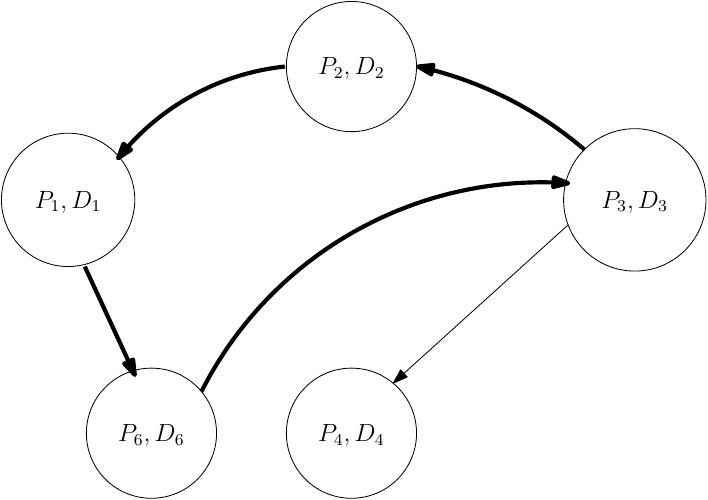}
    \caption{Second Iteration Compatibility Graph}
    \label{IKEA_compatibilityGraphWithMorePairs_secondIteration}
    \end{figure}
    
    \begin{figure}[H]
    \centering
    \includegraphics[height=3cm]{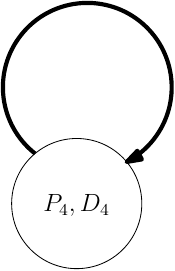}
    \caption{Third Iteration Compatibility Graph}
    \label{IKEA_compatibilityGraphWithMorePairs_thirdIteration}
    \end{figure}
    
    \item \textbf{Introduction of chains:} Let us assume an altruistic donor $A_1$, also known as Non directed donor(NDD). The details of which is shown in \cref{tab:altruisticDonorCharacteristicTable}. Let the societal distribution of $A_1$ be $sd_1$.
    
    \begin{table}[htbp]
    \caption{Details of Altruistic Donor}
    \label{tab:altruisticDonorCharacteristicTable}
    \vspace{2 mm}
    \small{\centerline{\begin{tabular}{c c c c c c} \hline
    Name  & ABO & HLA &Age &Size &PinCode \\ \hline
    $A_1$ & $A^+$  & A1,B8,DR17, &60 &11.5 &496001 \\
        &   &A10,B16,DR8    &   &   &   \\\hline
    \end{tabular}}}
    \end{table}
    
    \cref{IKEA_compatibilityGraphWithAltruisticDonor} describes the previous graph with an altruistic donor. Since the vertex $(P_5,D_5)$ is the most prioritized vertex, and according to the compatibility matrix in \cref{IKEA_compatibilityMatrixWithAltruisticDonor}, the altruistic donor $A_1$ is best suited for the vertex $(P_5,D_5)$. 
    
    \begin{figure}[H]
    \centering
    \includegraphics{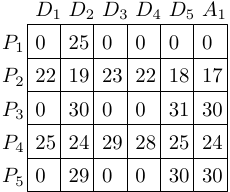}
    \caption{Compatibility Matrix With Altruistic Donor}
    \label{IKEA_compatibilityMatrixWithAltruisticDonor}
    \end{figure}
    
    \begin{figure}[H]
    \centering
    \includegraphics[width=8cm]{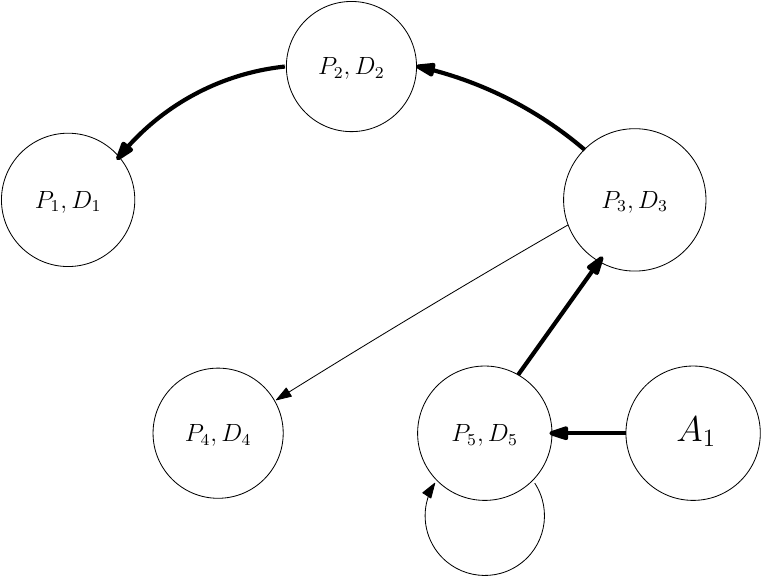}
    \caption{Compatibility Graph With NDD}
    \label{IKEA_compatibilityGraphWithAltruisticDonor}
    \end{figure}
    
    The chain as shown in \cref{IKEA_compatibilityGraphWithAltruisticDonor} which goes goes like $A_1 \longrightarrow (P_3,D_3) \longrightarrow (P_2,D_2) \longrightarrow (P_1,D_1)$, after getting executed leaves vertex $(P_4,D_4)$ unmatched in the first iteration. In the second iteration, $P_4$ receives kidney from the its donor $D_4$ itself as shown in \cref{IKEA_compatibilityGraphWithAltruisticDonorSecondIteration}.
    
    \begin{figure}[H]
    \centering
    \includegraphics[height=2.7cm]{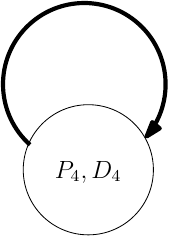}
    \caption{Second Iteration of IKEP with NDD}
    \label{IKEA_compatibilityGraphWithAltruisticDonorSecondIteration}
    \end{figure}
    
    And the best part about using an altruistic donors is that after completion of a chain, we are again left with a kidney of the donor of the last pair\footnote{The kidney of $D_1$, here.}. The donor can be a chain initiator for next immediate iteration or at any other point of time.
    
\end{itemize}

\section{Helicopter View Of IKEP} 

\begin{figure}[H]
    \centering
    \includegraphics[width=8cm]{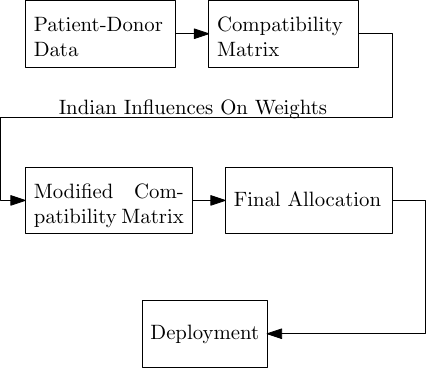}
    \caption{Overall architecture of IKEP}
    \label{IKEA_architecture}
    \end{figure}

\cref{IKEA_architecture} is a helicopter view of what we proposed in and as IKEP. IKEP can be broken down into smaller sections as, Receiving patient donor data, and then calculating compatibility matrix. After that the compatibility matrix is passed through Indian influence of weights, which gives us a new compatibility matrix i.e., Modified compatibility matrix. Now, this compatibility matrix is used for Final allocation. And at the end gets deployed.

\section{Future Scope}
\label{futureScope}
Since the presence of KEP in India is next to negligible, many problems need to be addressed and solved. Kidney exchange paradigm needs to be vigorously worked upon. Following are some of the immediate problems that needs to be addressed.
\subsection{Promoting truthful disclosure of pairs} 
Hospitals have a tendency, not to report all the pairs and keep some of them so that it can perform some matches internally. This strategy works against the idea of having a large exchange. And Ashlagi, I., et al. 2015\cite{27} did a study on how to make Hospitals report all there pairs and make the algorithm strategy-proof and came up with a mix and match algorithm. Hospitals in general have an incentive to match their own pairs if possible and register the hard to match pair for national kidney exchange, which in turn is leads to the accumulation of hard to match pairs.

Free riding is a major issue in KEPs, where hospitals are inclined towards preserving their easy to match pairs and disclosing their hard to match pairs. This tendency of hospitals lead to accumulation of hard to match pairs in the exchange pool.
.

\subsection{Dynamization of IKEP}

Dynamization of IKEP refers to the probabilistic calculation of parameters like $F^*$, probabilistic generation of a threshold value\footnote{Based on factors like arrival rates, tendency of patients to wait and many more}, exceeding which indicates algorithm run. The max length of cycles $L_{max}$ also needs to be determined according to the probability and statistics of patient-donor data. In other words, making IKEP an online algorithm.

\section{Conclusion}
\label{conclusion}
In this paper, we initially propose an algorithm(IKEPA) to run at a national level in India to accept the pairs and their preferences from the hospitals and in turn provide a considerably stable, individually rational and maximum cardinal exchanges. We propose some parameters which crucial to Indian perspective. But then found out that, particular implementation of Edge Filter($F^*$) is causing a huge amount of potential pairs to be rejected out, because of the, not up to the mark edges. Hence we come up with an enhanced form of IKEPA, named as E-IKEPA. It filters out those pairs, which have compatibility scores less than $F^*$ with their most preferred pair in the pool, and then run TTCA algorithm on the remaining pairs.

\section*{Appendix}
\appendix
\section{Class Definitions}
\label{classDefinitions}

Firstly, We describe the Vertex class in \cref{algo:ikep_vertexClass}, its member variables. The member variables are same as described in the \cref{prerequisites} and the ones not described are $ec$ which stores exclusion coefficient as described in \cref{exclusionCoefficient} and $rar$ stores the number of IKEPA runs after the pair rejected their kidney exchange offer.

\begin{algorithm}[H]
\caption{Vertex}\label{algo:ikep_vertexClass}
\begin{algorithmic}[1]
\Class{Vertex}
  \State $UID$ : \textsc{String}
  \State $Name$ : \textsc{String}
  \State $B$ : \textsc{String}
  \State $H$ : \textsc{List<String>}
  \State $Age$ : \textsc{Integer}
  \State $K$ : \textsc{String}
  \State $Pin$ : \textsc{Integer}
  \State $DName$ : \textsc{String}
  \State $DB$ : \textsc{String}
  \State $DH$ : \textsc{List<String>}
  \State $DAge$ : \textsc{Integer}
  \State $DK$ : \textsc{String}
  \State $DPin$ : \textsc{Integer}
  \State $PRA$ : \textsc{Integer}
  \State $Priority$ : \textsc{Integer}
  \State $isInitial$: \textsc{Boolean}
  \State $WTScore$ : \textsc{Integer}
  \State $Dsd$ : \textsc{String}
  \State $sdPref$ : \textsc{List<String>}
  \State $Eco$ : \textsc{Integer}
  \State $Dist$ : \textsc{Integer}
  \State $ec$ : \textsc{Integer}
  \State $rar$ : \textsc{Integer}
\EndClass
\end{algorithmic}
\end{algorithm}

We now describe the Graph class in \cref{algo:ikep_graphClass}, its member variables and member functions.

\begin{algorithm}[H]
\caption{Graph}\label{algo:ikep_graphClass}
\begin{algorithmic}[1]
\Class{Graph}
  \State $n$ : \textsc{Integer}
  \State $adj$ : \textsc{List<List<Vertex>>}
    \Procedure{Graph}{$n$}
    \State $this.n = n$
    \State $adj = $new ArrayList$<>(n)$
    \ForAll{$i \in \{0,1,\dots n\}$}
        \State $adj.$add(new LinkedList$<>())$
    \EndFor
    \EndProcedure
    \Procedure{addEdge}{$i,j$}\Comment{adds directed edge from i to j}
    \State $adj.$get$(i).$add$(j)$
    \EndProcedure
\EndClass
\end{algorithmic}
\end{algorithm}
\bibliographystyle{acm}
\bibliography{bibliography.bib}

\end{document}